\documentclass[]{tJOT2e}

\usepackage{graphicx} 
\usepackage{dcolumn}
\usepackage{bm}
\usepackage{amssymb}
\usepackage{amsmath}
\usepackage{wasysym}
\usepackage{color}
\usepackage{hyperref}
\usepackage{lscape}
\usepackage{rotating}
\usepackage{float}
\usepackage{flushend}
\usepackage{cancel}
\usepackage{hyperref}
\hypersetup{
    colorlinks,%
    citecolor=blue,%
    filecolor=blue,%
    linkcolor=blue,%
    urlcolor=blue
} 
\newcommand{\be} {\begin{equation}}
\newcommand{\ee} {\end{equation}}
\newcommand{\bea}{\begin{eqnarray}}
\newcommand{\eea}{\end{eqnarray}}

\begin{document}

\markboth{Mahendra K. Verma and Rohit Kumar}{Journal of Turbulence}

\articletype{}

\title{Dynamos at extreme magnetic Prandtl numbers: Insights from shell models}
\vskip 0.3 in

\author{Mahendra K. Verma$^{\rm a}$ and  Rohit Kumar$^{\rm a \ast}$ \thanks{$^\ast$Corresponding author. Email: rohitkumar.iitk@gmail.com
\vspace{6pt}} \\\vspace{6pt}  $^{\rm a}${\em{Department of Physics, Indian Institute of Technology, Kanpur 208016, India}}}


\maketitle

\begin{abstract}
We present an MHD shell model suitable for  computation of various energy fluxes of magnetohydrodynamic turbulence for very small and very large magnetic Prandtl numbers  $\mathrm{Pm}$; such computations are inaccessible to direct numerical simulations. For small  $\mathrm{Pm}$, we observe that both kinetic and magnetic energy spectra scale as $k^{-5/3}$ in the inertial range, but the dissipative magnetic energy scales as $k^{-11/3}\exp(-k/k_\eta)$.  Here, the kinetic energy at large length scale feeds the large-scale magnetic field that cascades to small-scale magnetic field, which gets dissipated by Joule heating.  The large-$\mathrm{Pm}$ dynamo has a similar behaviour except that the dissipative kinetic energy scales as $k^{-13/3}$.  For this case, the large-scale velocity field transfers energy to the large-scale magnetic field, which gets transferred to small-scale velocity and magnetic fields; the energy of the small-scale magnetic field also gets transferred to the small-scale velocity field, and the energy thus accumulated is dissipated by the viscous force.
\end{abstract}


\section{Introduction}

One of the celebrated problems in physics and astrophysics is the generation of magnetic field in planets, stars, and galaxies~\cite{Moffatt:book}.  The most popular  theory for this phenomenon is the dynamo mechanism in which the magnetic field is amplified due to the nonlinear energy transfer from the velocity field to the magnetic field.   The magnetic energy first grows and then saturates if the energy supply to the magnetic field could overcome the Joule dissipation.  In this paper we will report the energy spectra and fluxes during the steady-state using a popular model called {\em shell model}.

Some of the important parameters for dynamo are the magnetic Prandtl number $\mathrm{Pm} = \nu/\eta$ and the magnetic Reynolds number $\mathrm{Rm}=UL/\eta$, where $\nu$ is the kinematic viscosity, $\eta$ is the magnetic diffusivity, and $U,L$ are the velocity and length scales of the system.  A related quantity of importance is the kinetic Reynolds number $\mathrm{Re}=UL/\nu = \mathrm{Rm}/\mathrm{Pm}$.  The other important factors for dynamo are related to the rotation rate, temperature, viscous dissipation, geometry, etc.  
 
The sustained dynamo is possible only for some set of parameters.  For example, the energy inflow to the magnetic field at the large scale should, on the average, compensate the losses due to the Joule dissipation.  This condition and dimensional analysis yields $ U B /L \gtrapprox \eta B/L^2$ or $\mathrm{Rm} \gtrapprox 1$. However, numerical simulations and experimental studies indicate that the critical magnetic Reynolds number $\mathrm{Rm}_c > 10$~\cite{Yadav:EPL2010,Ponty:PRL2005,Monchaux:PRL2007}.   Kazantsev theory yields  a critical Rm of approximately 100. The magnetic Prandtl number also plays a major role in the dynamo mechanism.   Numerical and experimental studies show that the nature of small-$\mathrm{Pm}$ and large-$\mathrm{Pm}$ dynamos are very different.  The planetary and solar dynamos have typically small $\mathrm{Pm}$ (of the order of $10^{-5}$), while the interstellar dynamos have large $\mathrm{Pm}$ (of the order of $10^{11}$)~\cite{Brandenburg:PR2005,Plunian:PR2013}.   Unfortunately, direct numerical simulation (DNS) of small-$\mathrm{Pm}$ or large-$\mathrm{Pm}$ dynamos are very difficult due to large-scale separation between the velocity and magnetic dissipation regimes. At present, the largest and smallest $\mathrm{Pm}$ simulated so far are of the order of 100 and 0.01 respectively~\cite{Schekochihin:APJ2004,Haugen:PRE2004,Brandenburg:APJ2011}. Schekochihin et al.~\cite{Schekochihin:APJ2004} performed  dynamo simulations with large $\mathrm{Pm}$ ($25\leq \mathrm{Pm} \leq 2500$), while Haugen et al.~\cite{Haugen:PRE2004} and Brandenburg \cite{Brandenburg:APJ2011} simulated dynamos for $1/8 \leq \mathrm{Pm} \leq 30$ and  $0.01 \leq \mathrm{Pm} \leq 1$ respectively. Federrath {\it et al.}~\cite{Federrath:APJ2014} performed three-dimensional high resolution MHD simulation of supersonic turbulence with $\mathrm{Pm} = 0.1$---10 in which they observed dynamo for both high and low magnetic Prandtl numbers. They reported that the dynamo growth rate, and the ratio of the magnetic  energy and the kinetic energy at the saturation increase with $\mathrm{Pm}$. They also compared simulation results with theoretical predictions. Schober {\it et al.}~\cite{Schober:PRE2015} presented a theoretical model to study saturation of dynamo for $\mathrm{Pm} \gg 1$ and $\mathrm{Pm} \ll 1$; they observed that the saturation level of dynamo depends on the type of turbulence and the Prandtl number, and that the dynamos for $\mathrm{Pm} \gg 1$ are more efficient than those for $\mathrm{Pm} \ll 1$.

The magnetohydrodynamic (MHD) shell models are quite handy for simulating dynamos with very large and very small $\mathrm{Pm}$'s.  In a shell model, a single shell represents all the modes of a logarithmically binned shell.  Therefore, the number of variables required to simulate large-$\mathrm{Pm}$ and small-$\mathrm{Pm}$ dynamos using a shell model are much smaller than its DNS counterpart.  As an example, we can reach $\mathrm{Pm}=10^{-9}$ and $\mathrm{Pm}=10^{9}$ using shell models with $76$ shells.  There are a large number of MHD shell models~\cite{Gloaguen:PD1985,Biskamp:PRE1994,Brandenburg:PRD1996,Frick:PRE1998,Basu:PRL1998,Stepanov:JT2006,Lessinnes:PRE2009,Plunian:PR2013,Banerjee:PRL2013} that have yielded interesting results regarding the energy spectra and structure functions.  Some models yield Kolmogorov's energy spectrum $E(k) \sim k^{-5/3}$~\cite{Biskamp:PRE1994}, while others yield Kraichnan-Iroshnikov spectrum $E(k) \sim k^{-3/2}$ for the velocity and magnetic fields~\cite{Frick:PRE1998}.  

In the present paper we focus on the energy transfers of MHD turbulence using shell models.  A quantification of energy transfers is critical for understanding the growth and sustenance of the magnetic field, and it has been studied by several researchers~\cite{Moll:APJ2011,Kumar:EPL2013,Kumar:JT2015} using DNS, albeit at Prandtl numbers near unity. These energy transfer studies have also been performed by using theoretical tools~\cite{Verma:CS2002,Verma:PRAMANA2003a}. In this paper we perform similar analysis for extreme Prandtl numbers using a new shell model.  Earlier Stepanov and Plunian~\cite{Stepanov:JT2006}, Plunian {\em et al.}~\cite{Plunian:PR2013}, and Lessinnes {\em et al.}~\cite{Lessinnes:PRE2009} had derived formulas for the energy fluxes in their MHD shell models.  These models capture certain aspects of energy fluxes, but they fail to reproduce  all the results of DNS.  For example, in Lessinnes {\em et al.}'s model~\cite{Lessinnes:PRE2009}, most of the energy fluxes are in good qualitative agreement with the DNS results, but not the magnetic-to-magnetic flux ($\Pi^{B<}_{B>}$, defined in Appendix~\ref{appendix:MHD}).   For $\mathrm{Pm}=10^{-3}$, Lessinnes {\em et al.}'s model yields negative $\Pi^{B<}_{B>}$, but DNS by Dar {\it et al.}~\cite{Dar:PD2001},  Debliquy {\it et al.}~\cite{Debliquy:PP2005}, Alexakis {\it et al.}~\cite{Alexakis:PRE2005}, and Mininni {\it et al.}~\cite{Mininni:PRE2005} (also see Verma~\cite{Verma:PR2004}) yield positive $\Pi^{B<}_{B>}$. Note that all the above shell models yield the energy spectrum  correctly  due to the quadratic structures of the nonlinear terms.  Derivation of energy fluxes of MHD turbulence however requires more careful selection of the nonlinear terms, which is the objective of this paper.

In the present paper we construct a new shell model with structures suitable for studying energy transfers, as well as keeping all the conservation laws of MHD.  We observe that the nature of all the energy fluxes including $\Pi^{B<}_{B>}$ of this model are in good qualitative agreement with the DNS results~\cite{Dar:PD2001,Alexakis:PRE2005,Mininni:PRE2005,Debliquy:PP2005,Verma:PR2004}.  We study the properties of the energy fluxes during the steady state for small and large $\mathrm{Pm}$ using this new shell model.  We observe interesting correlations between the energy fluxes,  dissipation rates, and energy spectra.  A word of caution is that our shell model includes only the local nonlinear interactions among velocity and magnetic shells, hence the initial phases of the dynamo growth for large $\mathrm{Pm}$ is not captured properly by our shell model.  Inclusion of nonlocality in the shell model should be able to address this and related issues.  Also, we have ignored the magnetic and kinetic helicities in our shell model.  See Plunian {\em et al.}~\cite{Plunian:PR2013} for a detailed discussion on shell models with magnetic and kinetic helicities that modifies the energy transfers  significantly~\cite{Brandenburg:PR2005,Verma:PR2004,Plunian:PR2013}.
  
The structure of the paper is as follows: In section~\ref{sec:shell_mod_description}, we describe our MHD shell model. Section~\ref{sec:sim_detail} contains the simulation details and validation of the shell model.  In sections~\ref{sec:dyn_small_pm} and~\ref{sec:dyn_large_pm}  we present the results of our dynamo simulations with small and large $\mathrm{Pm}$, respectively. We summarize our findings in section~\ref{sec:conclu}.  Appendices A and B contain the derivation of energy fluxes for fluid and MHD shell models.


\section{Description of our MHD shell model}
\label{sec:shell_mod_description}

Our GOY-based shell model for dynamo is
\begin{eqnarray}
\frac{dU_{n}}{dt} & = & N_n[U,U] + N_n[B,B] - \nu k_n^2 U_n + F_n, \label{eq:shell_u} \\
\frac{dB_{n}}{dt} & = & N_n[U,B] + N_n[B,U] - \eta k_n^2 B_n, \label{eq:shell_b}
 \end{eqnarray}
where $U_n$ and $B_n$ represent respectively the velocity and the magnetic field variables in shell $n$, $F_n$ is the velocity forcing applied in shell $n$, $\nu$ is the kinematic viscosity,  $\eta$ is the magnetic diffusivity,  and $k_n = k_0 \lambda^n$ is the wavenumber of the $n$th shell with $\lambda = (\sqrt{5} + 1)/2$, the golden mean. The chosen $\lambda$ that satisfies $\lambda^2 - \lambda -1 = 0$.  The nonlinear terms $N_n[U,U]$, $N_n[B,B]$, $N_n[U,B]$, and $ N_n[B,U]$ correspond to respectively $-{\bf U}\cdot \nabla {\bf U}$,  ${\bf B}\cdot \nabla {\bf B}$, $-{\bf U}\cdot \nabla {\bf B}$, and ${\bf B}\cdot \nabla {\bf U}$, where ${\bf U,B}$ are the velocity and magnetic fields respectively,  of the MHD equations~\cite{Dar:PD2001,Verma:PR2004,Plunian:PR2013,Lessinnes:PRE2009}.  We derive the structures of the nonlinear terms of the shell model using the properties of the nonlinear transfers among the velocity and magnetic variables.  

The term $N_n[U,U]$ causes energy exchange among the velocity modes only, and it conserves total kinetic energy in this process, hence
 \begin{equation}
\Re \left(\sum_n U_n^* N_n[U,U] \right) = 0.
\end{equation}
The $B$ field plays no role in this transfer.  The  above constraint is same as that for the fluid shell model [Eq.~(\ref{eq:shell_u}) with $B=0$] in the inviscid limit.  Therefore, we choose same  $N_n[U,U] $ as the GOY fluid shell model (see e.g.~\cite{Gledzer:SPD1973}), which is
\begin{eqnarray}
N_n[U,U]  & = &  -i(a_{1}k_{n}U_{n+1}^{*}U_{n+2}^{*} + a_{2}k_{n-1}U_{n+1}^{*}U_{n-1}^{*} \nonumber \\ 
& & + a_{3}k_{n-2}U_{n-1}^{*}U_{n-2}^{*}), \label{eq:nonlin_uu} 
\end{eqnarray}
where $a_{1}, a_2$, and $a_3$ are constants. The conservation of the total kinetic energy, $E_u = (1/2) \sum_n \left[|U_{n}|^2 \right] $, in the inviscid limit with $B=0$ yields  
 \begin{equation}
a_1 + a_2 +a_3 = 0.
\label{eq:sum_a_i_0}
\end{equation}
In addition, we impose a condition that the kinetic helicity ($ H_{K}  = \sum_n (-1)^{n} |U_{n}|^{2} k_{n}$) is conserved for pure fluid case ($B=0$), which yields
\begin{equation}
a_{1}-\lambda a_{2}+\lambda^{2}a_{3}=0.
\label{eq:Hk_conservation}
\end{equation}

The nonlinear term $N_n[U,B] $, representing  $-{\bf U}\cdot \nabla {\bf B}$ of the MHD equation, facilitates energy exchange among the magnetic field variables with the velocity field variables acting as helper~\cite{Dar:PD2001,Verma:PR2004,Lessinnes:PRE2009}.   These transfers conserve the total magnetic energy, i.e.,
 \begin{equation}
\Re \left(\sum_n B_n^* N_n[U,B] \right) = 0.
\end{equation}
Keeping in mind the above constraint, we choose the following form for $N_n[U,B]$:
\begin{eqnarray}
N_n[U,B] & = & - i[k_{n}(d_{1}U_{n+1}^{*}B_{n+2}^{*} + d_{3}B_{n+1}^{*}U_{n+2}^{*}) \nonumber \\
& & + k_{n-1}(-d_{3}U_{n+1}^{*}B_{n-1}^{*} + d_{2}B_{n+1}^{*}U_{n-1}^{*}) \nonumber \\ 
& & + k_{n-2}(-d_{1}U_{n-1}^{*}B_{n-2}^{*} - d_{2}B_{n-1}^{*}U_{n-2}^{*})], \label{eq:nonlin_ub}
  \end{eqnarray}
where $d_{1}, d_2$, and $d_3$ are constants.  

The net loss of total energy (kinetic and magnetic) due to the aforementioned energy exchanges is zero for $\nu=\eta=0$~\cite{Verma:PR2004}.  Hence
\begin{equation}
\Re \left(\sum_n U_n^* N_n[B,B] + B_n^* N_n[B,U] \right) = 0, \label{eq:net_ub_bu_tr}
\end{equation}
where the term $U_n^* N_n[B,B]$, which corresponds to  $[{\bf B} \cdot \nabla {\bf B}] \cdot {\bf U}$ of the MHD equations, transfers energy from the magnetic energy to kinetic energy, while $B_n^* N_n[B,U]$, which  corresponds to $[{\bf B} \cdot \nabla {\bf U}] \cdot {\bf B}$, does just the opposite~\cite{Lessinnes:PRE2009}.
To satisfy Eq.~(\ref{eq:net_ub_bu_tr}), we choose the following forms of $N_n[B,B]$ and $N_n[B,U] $:
\begin{eqnarray}
N_n[B,B] & = & - 2i(b_{1}k_{n}B_{n+1}^{*}B_{n+2}^{*} + b_{2}k_{n-1}B_{n+1}^{*}B_{n-1}^{*} \nonumber \\ 
& & + b_{3}k_{n-2}B_{n-1}^{*}B_{n-2}^{*}), \label{eq:nonlin_bb} \\ 
N_n[B,U] & = & i[k_{n}(b_{2}U_{n+1}^{*}B_{n+2}^{*} + b_{3}B_{n+1}^{*}U_{n+2}^{*}) \nonumber \\ 
& & + k_{n-1}(b_{3}U_{n+1}^{*}B_{n-1}^{*}+b_{1}B_{n+1}^{*}U_{n-1}^{*}) \nonumber \\
&  & + k_{n-2}(b_{2}U_{n-1}^{*}B_{n-2}^{*} + b_{1}B_{n-1}^{*}U_{n-2}^{*})], \label{eq:nonlin_bu} 
 \end{eqnarray}
where $b_{1}, b_2$, and $b_3$ are constants.  Note that above choices of the nonlinear terms satisfy the conservation of total energy of the MHD shell model
\begin{equation}
E = E_u + E_b  =   \frac{1}{2} \sum_n \left[|U_{n}|^2 + |B_{n}|^2 \right] 
\end{equation}
where $E_u = (1/2) \sum_n \left[|U_{n}|^2 \right] $ and $E_b = (1/2) \sum_n \left[|B_{n}|^2 \right] $ are the total kinetic energy and the total magnetic energy respectively. Note that we need to impose only the constraint of Eq.~(\ref{eq:sum_a_i_0}) for the above conservation law.  The nonlinear terms $N_n[B,B]$, $N_n[U,B]$, and $ N_n[B,U]$ conserve the total energy automatically. Also, Eq.~(\ref{eq:sum_a_i_0}) insures that the kinetic energy is conserved for pure fluid case ($B=0$) when $\nu=0$.

The four nonlinear terms contain 9 undetermined constants $a_1, a_2, a_3$, $b_1, b_2, b_3, d_1,d_2$, and $d_3$.  We determine their values using constraint equations, two of which are  Eqs.~(\ref{eq:sum_a_i_0}) and (\ref{eq:Hk_conservation}). We also use the other two conservations laws, which are  the conservation of the total cross helicity $H_c$ and the total magnetic helicity $H_M$ when $\nu=\eta=0$ and $F_n=0$.  These quadratic conserved quantities are defined as
\begin{eqnarray}
 H_{c} & = & \frac{1}{2}\Re  \sum_n U_{n}B_{n}^{*} \\
 H_{M} & = & \sum_n (-1)^{n}|B_{n}|^{2}/k_{n},
 \end{eqnarray}
The conservation of cross helicity yields
\begin{eqnarray}
b_{1}+b_{2}+b_{3} & = & 0,\label{eq:bd-1}\\
a_{1}-b_{3}-d_{3}-b_{2}-d_{1} & = & 0\label{eq:bd-2}\\
a_{2}+d_{3}-b_{3}-b_{1}-d_{2} & = & 0\label{eq:bd-3}\\
a_{3}-b_{2}+d_{1}-b_{1}+d_{2} & = & 0,\label{eq:bd-4}
\end{eqnarray}
while the conservation of magnetic helicity  yields
\begin{eqnarray}
\lambda^{-2}(-d_{1}-b_{2})-(-d_{1}+b_{2}) & = & 0\label{eq:bd-5}\\
\lambda^{-2}(-d_{2}-b_{1})+\lambda^{-1}(-d_{2}+b_{1}) & = & 0\label{eq:bd-6}\\
\lambda^{-1}(b_{3}+d_{3})-(-d_{3}+b_{3}) & = & 0.\label{eq:bd-7}
\end{eqnarray}

Thus we have nine constraints [Eqs.~(\ref{eq:sum_a_i_0},  \ref{eq:Hk_conservation}, \ref{eq:bd-1}-\ref{eq:bd-7}], and nine unknowns ($a_1, a_2, a_3, b_1, b_2, b_3, d_1, d_2, d_3$).  However, the determinant of the matrix formed by these equations is zero.  Hence,  the solution of the above equations is not unique.  One of the solutions that satisfies all the above constraints is given below: 
\begin{eqnarray}
a_{1} &= & \lambda, a_{2} = 1-\lambda, a_{3} = -1,   \nonumber \\
b_{1} & = & \lambda, b_{2}  =  1+\frac{\lambda}{2}, b_{3} = -1-\frac{3\lambda}{2}, \nonumber \\
  d_{1}&  = & \frac{5\lambda}{2}, d_{2}=  -\lambda +2, d_{3} = -\frac{\lambda}{2}.
\end{eqnarray} 
It can be shown that ($\delta a_1,  \delta a_2, \delta a_3, \delta b_1, \delta b_2, \delta b_3, \delta d_1,\delta  d_2, \delta d_3$), where $\delta$ is  a constant, also satisfy the equations. 
We remark that some free parameters are chosen in almost all the shell models.   Biferale~\cite{Biferale:ARFM2003} and Ditlevsen~\cite{Ditlevsen:book} attribute  the above arbitrariness of the parameter to the freedom for the choice of the time scale.  The above factor $\delta$ is the arbitrary prefactor. Note that  $\lambda$ (the golden mean) in the above equations satisfies the equation  $\lambda^2 - \lambda -1 = 0$. 

The aforementioned  nonlinear terms $N_n[U,U]$, $N_n[B,B]$, $N_n[U,B]$, and $N_n[B,U]$ facilitate energy transfers from velocity-to-velocity ($U2U$), magnetic-to-velocity ($B2U$), magnetic-to-magnetic ($B2B$), and velocity-to-magnetic ($U2B$) respectively, and they induce energy fluxes of MHD turbulence that play a critical role in the dynamo mechanism.    The energy flux $\Pi^{X<}_{Y>}(K)$ is the rate of transfer of the energy of the field $X$  from the shells inside the sphere of radius $K$ to the field $Y$ outside the sphere.   The above flux can be written in terms of the energy transfer formulas derived in Appendix~\ref{appendix:MHD}:
\begin{eqnarray}
\Pi^{X<}_{Y>}(K) & = & \sum_{m\le K} \sum_{n>K} \sum_{p}S^{YX}(n|m|p) \label{eq:flux_X-to-Y}.  
\end{eqnarray}
In particular
\begin{eqnarray}
\Pi^{U<}_{U>}(K) & = & \sum_{m\le K} \sum_{n>K} \sum_{p}S^{UU}(n|m|p) \label{eq:fluxUU} \\ 
\Pi^{B<}_{B>}(K) & = & \sum_{m\le K} \sum_{n>K} \sum_{p}S^{BB}(n|m|p) \label{eq:fluxBB} \\ 
\Pi^{U<}_{B>}(K) & = & \sum_{m\le K} \sum_{n>K} \sum_{p}S^{BU}(n|m|p) \label{eq:fluxUB} \\ 
\Pi^{B<}_{U>}(K) & = & \sum_{m\le K} \sum_{n>K} \sum_{p}S^{UB}(n|m|p). \label{eq:fluxBU} 
\end{eqnarray}
The energy flux $\Pi^{U<}_{B<}(K)$  is the energy flux from the $U$-shells inside the sphere to the $B$-shells inside the sphere, while $\Pi^{U>}_{B>}(K)$ is the corresponding energy flux for the shells outside the sphere.  We refer to Appendix~\ref{appendix:MHD} for a detailed derivation of energy fluxes.  

In the following section we also report the kinetic and magnetic energy spectra which are defined as
\begin{eqnarray}
E_u(k) & = &  \frac{1}{2} \frac{|U_n|^2}{k_n} \label{eq:Eu_k} \\ 
E_b(k) & = &  \frac{1}{2} \frac{|B_n|^2}{k_n}  \label{eq:Eb_k} 
\end{eqnarray}
In the inertial range, we observe that $|U_n|^2 \sim |B_n|^2 \sim k_n^{-2/3}$, hence the energy spectra appear to scale as $k_n^{-5/3}$, which is the Kolmogorov's spectrum.

We  point out that our shell model differs from earlier ones due to the aforementioned  structures of energy transfers.   Biskamp's shell model~\cite{Biskamp:PRE1994} uses Els\"{a}sser variables.  The form of the nonlinear terms of Stepanov and Plunian's shell model is very different from ours (see Eqs.~(1-3) of ~\cite{Stepanov:JT2006}).  In the shell model of Frick {\it et al}.~\cite{Frick:NJP2007}, $d B_n/dt$ does not involve $U_n$ at all, while the shell model of Plunian and Stepanov~\cite{Plunian:NJP2007} includes nonlocal terms. In the shell model of Stepanov and Plunian~\cite{Stepanov:JT2006}, Lessinnes {\em et al.}~\cite{Lessinnes:PRE2009}, and Stepanov and Plunian~\cite{Stepanov:APJ2008}, the structures of $N_n[B,B]$ and $N_n[U,U]$ are the same, and so are those of $N_n[U,B]$ and $N_n[B,U]$.  As a result, these models do not yield correct  magnetic-to-magnetic energy transfers ($B2B$) due to $N_n[U,B]$, as well as magnetic-to-velocity ($B2U$) and velocity-to-magnetic ($U2B$) transfers arising due to $N_n[B,B]$ and $N_n[B,U]$.    This is the reason for the negative $B2B$ flux arising in these models.  As described above, our model corrects these deficiencies in the previous shell models and it yields correct energy transfers.  We also remark that earlier shell models reproduce the kinetic and magnetic energy spectra quite correctly since they depend on the quadratic structure of the shell model.  The energy transfers however require more sophisticated structuring of the nonlinear terms.

In the next section, we provide simulation details.

\section{Simulation details and validation for $\mathrm{Pm} =1$}
\label{sec:sim_detail}

In our shell model simulations, we divide the wavenumber space into $36$ or $76$ logarithmically binned shells. The large number of shells have been used for simulating dynamos with $\mathrm{Pm} =10^{-9}$ and $10^{9}$.  We use the forcing scheme of Stepanov and Plunian~\cite{Stepanov:JT2006} in which the forcing is applied to three neighbouring shells $n_f$, $n_f + 1$, and $n_f + 2$ as $F_{n_f+j} =f_j e^{i \phi_j}$ ($j=0,1,2$).  Here $f_j$'s are real positive numbers derived in Stepanov and Plunian~\cite{Stepanov:JT2006}, and $\phi_j \in [0, 2\pi]$ are random phases. In our simulations, we employ external random forcing to the  velocity shells 3, 4, and 5 such that the kinetic energy supply rate is maintained at a constant value ($\epsilon = 1$), and the normalized kinetic and magnetic helicities as well the normalized cross helicity are relatively small.

We first initiate a pure fluid simulation with a random initial condition for the velocity field and run the simulation till it reaches a statistically steady state. For the initial condition  for a dynamo simulation, we take the above steady fluid state as the initial velocity configuration and  a small seed magnetic field at shells 1 and 2.  The dynamo simulation is carried out till it reaches a steady state. For the time integration, we employ Runge-Kutta fourth order (RK4) scheme with a fixed $\Delta t$. The choice of a fixed $\Delta t$ helps during the kinematic growth phase where dynamic $\Delta t$ varies widely. In our forced MHD simulations, the kinetic and magnetic energies saturate at $t \approx 10$, but we carry out the simulations for a much longer time.  Here the unit of time is the eddy turnover time.  For the computation of the energy spectrum and energy fluxes, we average these quantities (in steady state) for long time intervals. 

We perform our simulations for 6 sets of parameters listed in Table~\ref{table:sim_details}.  The magnetic Prandtl numbers for these runs are $\mathrm{Pm} =1, 10^{-3}$, $10^{3}$, $10^{-9}$ and $10^9$, thus we cover very small to very large $\mathrm{Pm}$'s.  Note that the steady state kinetic Reynolds number and magnetic Reynolds number are quite large, hence our runs are in the turbulent regime.  In our shell model simulations, the total kinetic and total magnetic energies fluctuate considerably with time.   Therefore, we compute the time-averaged values of the total kinetic energy ($E_u$), the total magnetic energy ($E_b$), and their ratio, and present them in Table~\ref{table:en_diss_hel}.   In the forced dynamo simulations, the fluctuations in $E_u/E_b$ is of the order of unity, whereas in the decaying simulation, it is of the order of $10^{-2}$.  Stepanov and Plunian~\cite{Stepanov:JT2006}   also observed similar fluctuations in kinetic and magnetic energies in their shell model of dynamo.

We compute the kinetic and magnetic helicities.  The normalized kinetic and magnetic  helicities, as well as the normalized cross helicity are small (less than a quarter), hence all our runs are nonhelical (see Table~\ref{table:en_diss_hel}). In Table~\ref{table:en_diss_hel} we also exhibit the kinetic and magnetic energy dissipation rates as well as their ratio.  We observe that the  ratio of the dissipation rates increases with $\mathrm{Pm}$, from very small values for $\mathrm{Pm} \ll 1$ to very large values for $\mathrm{Pm} \gg 1$, similar to that observed by Brandenburg~\cite{Brandenburg:APJ2014}.

\begin{table}[htbp]
\caption{Parameters of shell model simulations (D = decaying, F= forced): number of shells ($n$), kinematic viscosity ($\nu$), magnetic diffusivity ($\eta$), magnetic Prandtl number ($\mathrm{Pm} =\nu/\eta$), kinetic Reynolds number ($\mathrm{Re} =UL/\nu$), and magnetic Reynolds number ($\mathrm{Rm} =UL/\eta$).} 
\vspace{0.2cm}
\centering
\begin{tabular}{c c c c c c c}
\hline 
\hline  
Sim & $n$ & $\nu$ & $\eta$ & $\mathrm{Pm}$ & $\mathrm{Re}$ & $\mathrm{Rm}$  \\ [1ex] 
\hline
D & 36  & $10^{-6}$ & $10^{-6}$ & $1$ & $3.23 \times 10^5$ & $3.23 \times 10^5$  \rule{0pt}{3ex}  \\

F1 & 36 & $10^{-6}$ & $10^{-6}$ & $1$ & $9.35 \times 10^6$ & $9.35 \times 10^6$   \\ 

F2 & 36 & $10^{-9}$ & $10^{-6}$ & $10^{-3}$ & $9.21 \times 10^{9}$ & $9.21 \times 10^6$  \\

F3 & 36 & $10^{-6}$ & $10^{-9}$ & $10^3$ & $9.68 \times 10^6$ & $9.68 \times 10^{9}$  \\

F4 & 76 & $10^{-13}$ & $10^{-4}$ & $10^{-9}$ & $9.16 \times 10^{13}$ & $9.16 \times 10^4$  \\

F5 & 76 & $10^{-6}$ & $10^{-15}$ & $10^9$ & $8.94 \times 10^6$ & $8.94 \times 10^{15}$  \\

\hline
\hline
\end{tabular}
\label{table:sim_details} 
\end{table}

\begin{table}[htbp]
\caption{Parameters of shell model simulations: magnetic Prandtl number, time-averaged values of  total kinetic energy ($\left\langle E_u \right\rangle$), total magnetic energy ($\left\langle E_b \right\rangle$),  $\left\langle E_u \right\rangle / \left\langle E_b \right\rangle$,    total kinetic energy dissipation rate ($\left\langle \epsilon_\nu \right\rangle$), total magnetic energy dissipation  rate ($\left\langle \epsilon_\eta \right\rangle$),  $\left\langle \epsilon_\nu \right\rangle / \left\langle \epsilon_\eta \right\rangle$,  normalized kinetic helicity ($\left\langle h_K \right\rangle = \left\langle (\sum_n (-1)^{n}|U_{n}|^{2} k_{n}) / (\sum_n |U_{n}|^{2} k_{n}) \right\rangle$), normalized magnetic helicity ($\left\langle h_M \right\rangle = \left\langle (\sum_n (-1)^{n}|B_{n}|^{2} / k_{n}) / (\sum_n |B_{n}|^{2} / k_{n}) \right\rangle$), and normalized cross helicity ($\left\langle h_c \right\rangle = \left\langle (\Re  \sum_n 2 U_{n}B_{n}^{*}) / \sum_n (|U_{n}|^{2} + |B_{n}|^{2}) \right\rangle)$.} 
\vspace{0.2cm}
\centering
\begin{tabular}{c c c c c c c c c c c}
\hline 
\hline  
Sim &  $\mathrm{Pm}$ & $\left\langle E_u \right\rangle$ & $\left\langle E_b \right\rangle$ & $\frac{\left\langle E_u \right\rangle}{\left\langle E_b \right\rangle}$ & $\left\langle \epsilon_\nu \right\rangle$ & $\left\langle \epsilon_\eta \right\rangle$ & $\frac{\left\langle \epsilon_\nu \right\rangle}{\left\langle \epsilon_\eta \right\rangle}$ & $\left\langle h_K \right\rangle$ & $\left\langle h_M \right\rangle$ & $\left\langle h_c \right\rangle$ \\ [1ex] 
\hline

F1 & $1$ & $1.14$ & $0.91$ & $1.25$ & $0.39$ & $0.58$ & $0.67$ & $0.27$ & $0.15$ & $-0.02$  \rule{0pt}{3ex} \\

F2 & $10^{-3}$ & $1.10$ & $0.88$ & $1.25$ & $0.08$ & $0.90$ & $0.09$ & $0.20$ & $0.16$ & $-0.02$  \\

F3 & $10^3$ & $1.22$ & $0.90$ & $1.36$ & $0.94$ & $0.04$ & $23.50$ & $0.23$ & $0.15$  & $-0.02$ \\

F4 & $10^{-9}$ & $1.05$ & $0.76$ & $1.38$ & $0.06$ & $0.91$ & $0.07$ & $0.07$ & $0.15$ & $0.04$ \\

F5 & $10^9$ & $1.03$ & $0.93$ & $1.11$ & $0.96$ & $\rightarrow 0$ & $\rightarrow \infty$ & $0.23$ & $0.19$ & $-0.04$ \\

\hline
\hline
\end{tabular}
\label{table:en_diss_hel} 
\end{table}

\begin{figure}[htbp]
\centering
\includegraphics[scale=0.48]{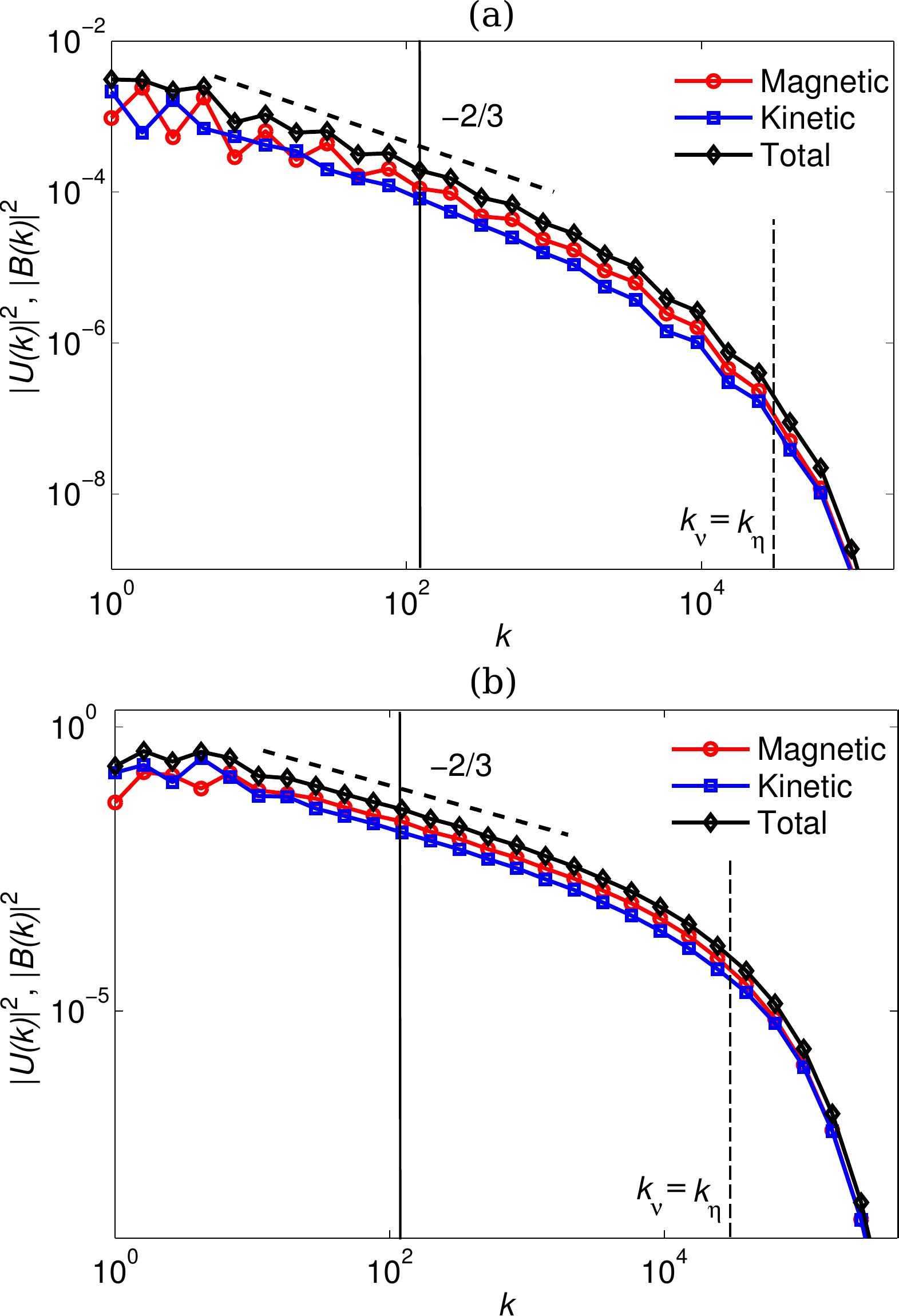}
\caption{For $\mathrm{Pm} = 1$: Time-averaged kinetic energy ($|U(k)|^2$), magnetic energy ($|B(k)|^2$), and total energy ($|U(k)|^2 + |B(k)|^2$) in spectral space for (a) decaying simulation and (b) forced simulation.  See Table~\ref{table:sim_details} for details. Fig.~\ref{fig:energy_flux_pm_1_schem} exhibits the energy fluxes at $K$ corresponding to the vertical solid line. Here $k_\nu = (\epsilon/\nu^3)^{1/4}$ and $k_\eta = (\epsilon/\eta^3)^{1/4}$.}
\label{fig:energy_spec_pm_1}
\end{figure}

For validation of our flux formulas, we performed decaying and forced shell model simulations for $\mathrm{Pm}=1$ with 36 shells.  In Fig.~\ref{fig:energy_spec_pm_1} we plot the kinetic energy spectrum $|U(k)|^2$, the magnetic energy spectrum $|B(k)|^2$, and the total energy spectrum $|U(k)|^2$ + $|B(k)|^2$, all of which exhibit an approximate $k^{-2/3}$ spectrum in correspondence with the Kolmogorov's spectrum of $E_u(k) \sim E_b(k) \sim k^{-5/3}$ [See Eqs.~(\ref{eq:Eu_k}, \ref{eq:Eb_k})].   The aforementioned results have been observed for a large number of shell models~\cite{Frick:PRE1998,Stepanov:JT2006,Plunian:NJP2007,Lessinnes:PRE2009}.  We compute the shell model spectral exponents using linear regression on the log-log plots. The exponents with the corresponding errors for various simulations are listed in Table~\ref{table:exp_err}. Note that the dissipation of $E_u(k)$ starts near Kolmogorov wavenumber $k = k_\nu = (\epsilon/\nu^3)^{1/4}$.  The corresponding dissipative wavenumber for $E_b(k)$ is $ k_\eta = (\epsilon/\eta^3)^{1/4}$.

\begin{table}[htbp]
\caption{ Exponents of the kinetic  and magnetic energy spectra with corresponding errors.}
\vspace{0.2cm}
\centering
\begin{tabular}{c c c c}
\hline 
\hline  
Sim & $\mathrm{Pm}$ & KE exponent & ME exponent \\ [1ex] 
\hline
D  & $1$ & $-0.67 \pm 0.07$ & $-0.80 \pm 0.40$  \rule{0pt}{3ex}  \\

F1 & $1$ & $-0.67 \pm 0.03$ & $-0.68 \pm 0.05$   \\ 

F2 & $10^{-3}$ & $-0.68 \pm 0.04$ & $-0.67 \pm 0.07$  \\

F3 & $10^3$ & $-0.67 \pm 0.02$ & $-0.65 \pm 0.04$  \\

F4 & $10^{-9}$ & $-0.65 \pm 0.07$ & $-0.70 \pm 0.20$  \\

F5 & $10^9$ & $-0.67 \pm 0.01$ & $-0.70 \pm 0.20$  \\

\hline
\hline
\end{tabular}
\label{table:exp_err} 
\end{table}

\begin{figure}[htbp]
\centering
\includegraphics[scale=0.45]{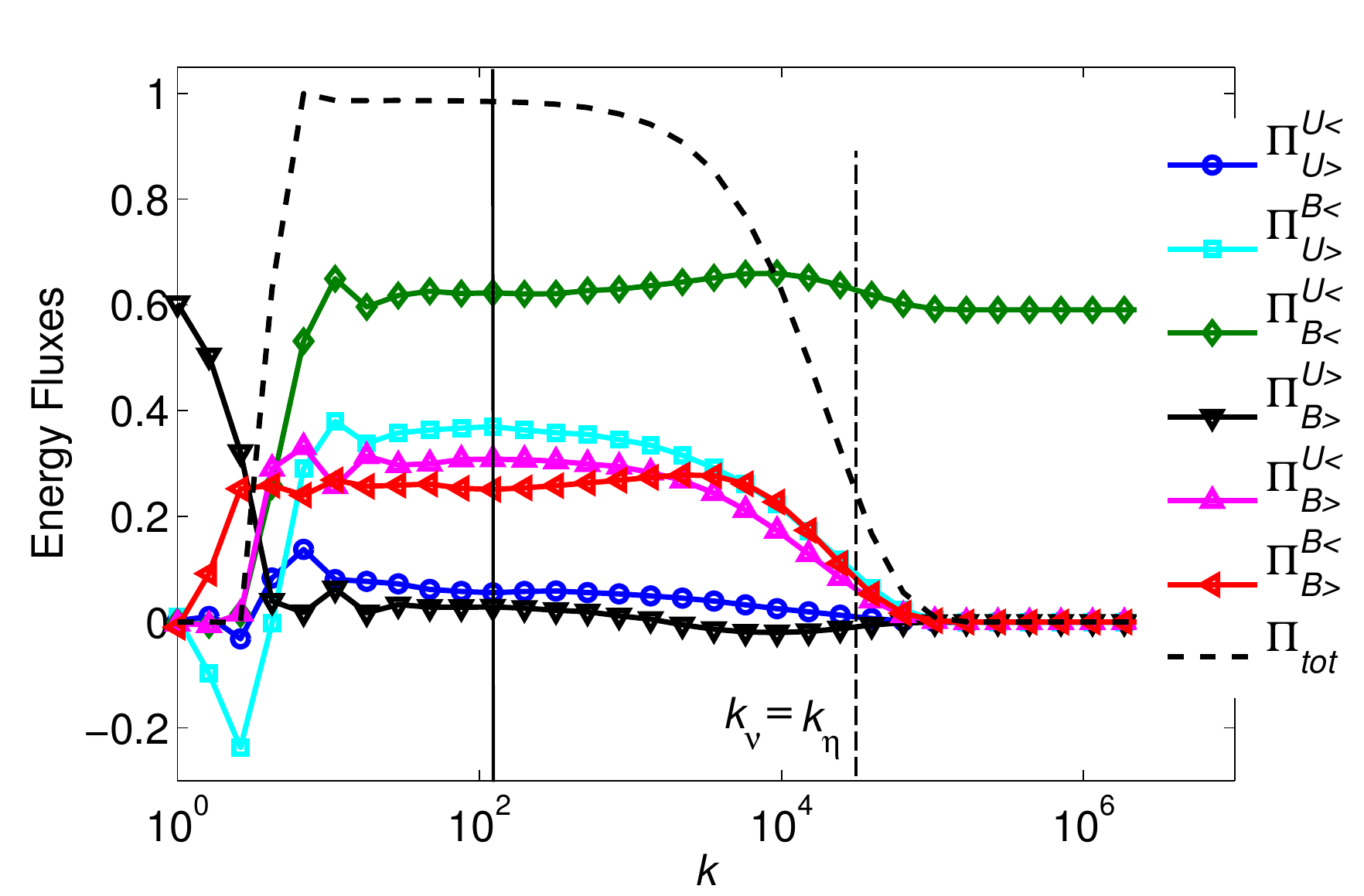}
\caption{For forced dynamo simulation with $\mathrm{Pm} = 1$: Time-averaged steady state energy fluxes: $\Pi^{U<}_{U>}$, $\Pi^{B<}_{U>}$, $\Pi^{U<}_{B<}$, $\Pi^{U>}_{B>}$, $\Pi^{U<}_{B>}$, $\Pi^{B<}_{B>}$, and $\Pi_{tot} = \Pi^{U<}_{U>} + \Pi^{U<}_{B>} + \Pi^{B<}_{U>} + \Pi^{B<}_{B>}$.  Fig.~\ref{fig:energy_flux_pm_1_schem} exhibits the energy fluxes at $K$ corresponding to the vertical solid line.}
\label{fig:energy_flux_pm_1}
\end{figure}

\begin{figure}[htbp]
\centering
\includegraphics[scale=0.27]{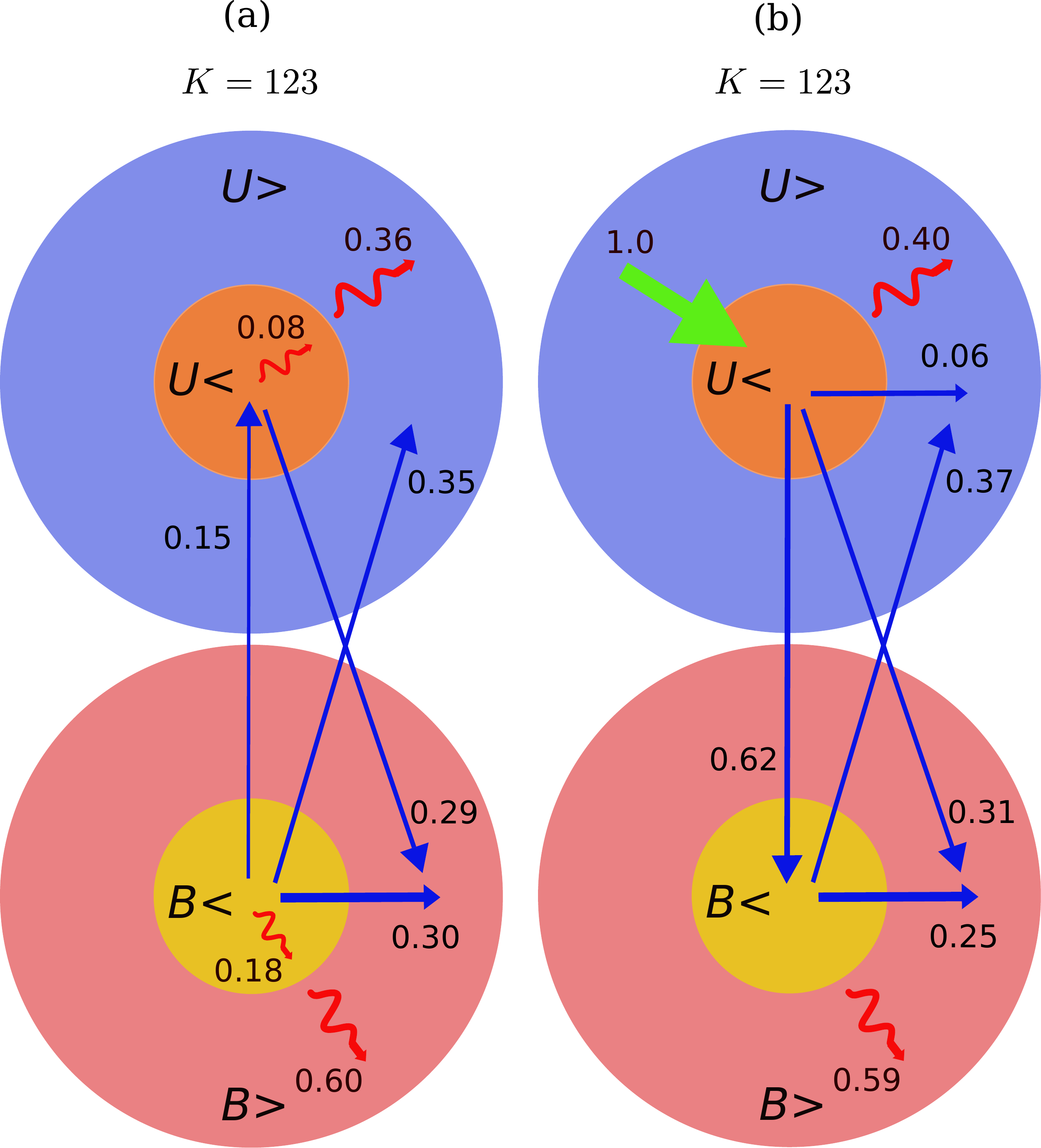}
\caption{For dynamo simulation with $\mathrm{Pm} = 1$, schematic diagrams of the time-averaged energy fluxes and dissipation rates for the wavenumber sphere of radius $K = 123$: (a)   decaying simulation, (b) forced simulation. The green arrow indicates kinetic energy supply rate, whereas the wavy red arrows indicate kinetic energy and magnetic energy dissipation rates. The uncertainty in the values are approximately 0.02. }
\label{fig:energy_flux_pm_1_schem}
\end{figure}

The primary advantage of our shell model is its ability to compute the energy fluxes of MHD turbulence. We compute the fluxes for the decaying as well as the forced shell models by averaging over a long time.  For the decaying run, the averaging has been performed when the  variations in the energy are reasonably small.  Various energy fluxes for the forced shell model simulation are shown in Fig.~\ref{fig:energy_flux_pm_1} that shows constant values for these fluxes in the inertial range.  In Fig.~\ref{fig:energy_flux_pm_1_schem}(a,b)  we exhibit the constant values of the fluxes for the wavenumber sphere of radius $K = 123$ for the decaying and forced simulations.  In Table~\ref{table:flux_dec_pm_1} we  list these values along with the DNS results of  Dar {\it et al.}~\cite{Dar:PD2001} and Debliquy {\it et al.}~\cite{Debliquy:PP2005}. 

Some of the important features of our flux results are:
\begin{itemize}
\item The Alfv\'{e}n ratio $r_A = E_u/E_b$ is approximately 0.5 for the decaying simulation towards its later phase.  However it is approximately 1.5 for the forced run during the steady state.  Note that  DNS yield $r_A \approx 0.5$ for both decaying (at a later stage) and forced runs.

\item For the forced run, $\Pi^{U<}_{B<}$ is the most dominant flux indicating a strong energy transfer from the large-scale velocity field to the large-scale magnetic field.  The energy received by the large-scale (small $k$) magnetic field is transferred to the small-scale (large $k$) magnetic field by a {\em forward cascade} of magnetic energy ($\Pi^{B<}_{B>} > 0$). This result of ours is in good qualitative agreement with that of Dar {\it et al.}~\cite{Dar:PD2001}.  

\item For the forced run, the energy fluxes $\Pi^{U<}_{U>}$, $\Pi^{U<}_{B>}$, $\Pi^{B<}_{U>}$, $\Pi^{U>}_{B>}$ are all positive.  The visible differences between the DNS and shell model may be because Dar {\it et al.}~\cite{Dar:PD2001} simulated  two-dimensional MHD turbulence.

\item For $r_A \approx 0.5$,  $\Pi^{U<}_{B<} < 0$ implying that the large-scale magnetic field transfers energy to the large-scale velocity field.  This result is in agreement with the DNS results of Debliquy {\it et al.}~\cite{Debliquy:PP2005}, who argue that  in decaying simulations, $\Pi^{U<}_{B<} < 0$ for $E_b > E_u$ and vice versa.  Our shell model is consistent with the above observations of Debliquy {\it et al.}

\item In both decaying and forced shell model simulations, the  $\Pi^{B<}_{B>} > 0$, which is consistent with the DNS results.  The Lessinnes {\it et al.}'s shell model~\cite{Lessinnes:PRE2009} however yield $\Pi^{B<}_{B>} < 0$. Hence, our shell model is a better candidate for computing energy fluxes of MHD turbulence than that of Lessinnes {\it et al.}  We also remark that the Stepanov and Plunian's~\cite{Stepanov:JT2006} formula for $\Pi^{B<}_{B>} $ has certain ambiguities. 

\item The errors in the computed energy fluxes are 2\%. Hence the uncertainty in the energy balance, e.g., the difference between the sum of the dissipation rates and the energy supply rate, is approximately 2\%. 
 
\end{itemize}

The above results indicate that our shell model is a good candidate for studying energy transfers in MHD turbulence. 
  
\begin{table}[htbp]
\caption{For decaying and forced MHD simulation with $\mathrm{Pm} =1$: Time-averaged energy fluxes, viscous dissipation rate ($\epsilon_\nu$), and Joule dissipation rate ($\epsilon_\eta$) for $K =123$ in our shell model.  We also list these quantities for the decaying 3D dynamo simulation by Debliquy {\it et al.}~\cite{Debliquy:PP2005}, and for the forced 2D dynamo simulation by Dar {\it et al.}~\cite{Dar:PD2001}. The uncertainty in the values are approximately 0.02. Here $-$ represents unavailable values.} 
\vspace{0.2cm}
\centering
\begin{tabular}{c c c c c}
\hline 
\hline  
Flux & $r_A =0.60$ &  $r_A = 0.50 $ & $r_A =0.50$ & $r_A =1.50$\\  
	 & (Deb, DNS) & (Dar, DNS)  & (shell model) & (shell model) \\
  	 & (Decaying) & 	 (Forced)  & (Decaying) 	  & (Forced)\\ 
  	 & ($K \approx 18$)	  &  ($K = 20$)  & ($K = 123$)	  & ($K = 123$)	\\ [1ex]
\hline 
$\Pi^{U<}_{U>}$ & $0.07$ & $-0.13$ & $0.01$ & $0.06$ \rule{0pt}{3ex} \\

$\Pi^{U<}_{B>}$ & $0.49$ & $0.68$ & $0.29$ & $0.31$ \rule{0pt}{4ex} \\

$\Pi^{B<}_{U>}$ & $0.13$ & $-0.09$ & $0.35$ & $0.37$ \rule{0pt}{4ex} \\

$\Pi^{B<}_{B>}$ & $0.36$ & $0.47$ & $0.30$ & $0.25$ \rule{0pt}{4ex} \\

$\Pi^{U<}_{B<}$ & $-0.02$ & $0.37$ & $-0.15$ & $0.62$ \rule{0pt}{4ex} \\

$\Pi^{U>}_{B>}$ & $0.22$ & $-0.42$ & $0.01$ & $0.03$ \rule{0pt}{4ex} \\

$\epsilon_\nu(U \textless)$ & $-$ & $-$ & $0.08$ & $0.00$ \rule{0pt}{4ex} \\

$\epsilon_\eta(B \textless)$ & $-$ & $-$ & $0.18$ & $0.00$ \rule{0pt}{3ex} \\

$\epsilon_\nu(U \textgreater)$ & $-$ & $0.39$ & $0.36$ & $0.40$ \rule{0pt}{3ex} \\

$\epsilon_\eta(B \textgreater)$ & $-$ & $0.55$ & $0.60$ & $0.59$ \rule{0pt}{3ex} \\
[0.5ex]
\hline
\hline
\end{tabular}
\label{table:flux_dec_pm_1} 
\end{table}

The dynamo mechanism is typically studied for small $\mathrm{Pm}$, mimicking the stellar and planetary dynamos, and for large $\mathrm{Pm}$ corresponding to the astrophysical plasmas. In the following sections we will describe the energy spectra and fluxes for such dynamos.

\section{Dynamo with small Prandtl numbers}
\label{sec:dyn_small_pm}

We performed dynamo simulations for $\mathrm{Pm}=10^{-3}$ and $10^{-9}$, which are completely inaccessible to the DNS at present.  The numbers of shells used for these runs are $36$ and $76$ respectively.  First we simulate the fluid shell model ($B=0$) till it reaches the steady state.  After this we start our dynamo simulations with the above steady-state velocity field and a random seed magnetic field.  The dynamo is forced at low wavenumbers (shells $3$-$5$) using the method of Stepanov and Plunian (described in Sec.~\ref{sec:sim_detail}).

\begin{figure}[htbp]
\centering
\includegraphics[scale=0.55]{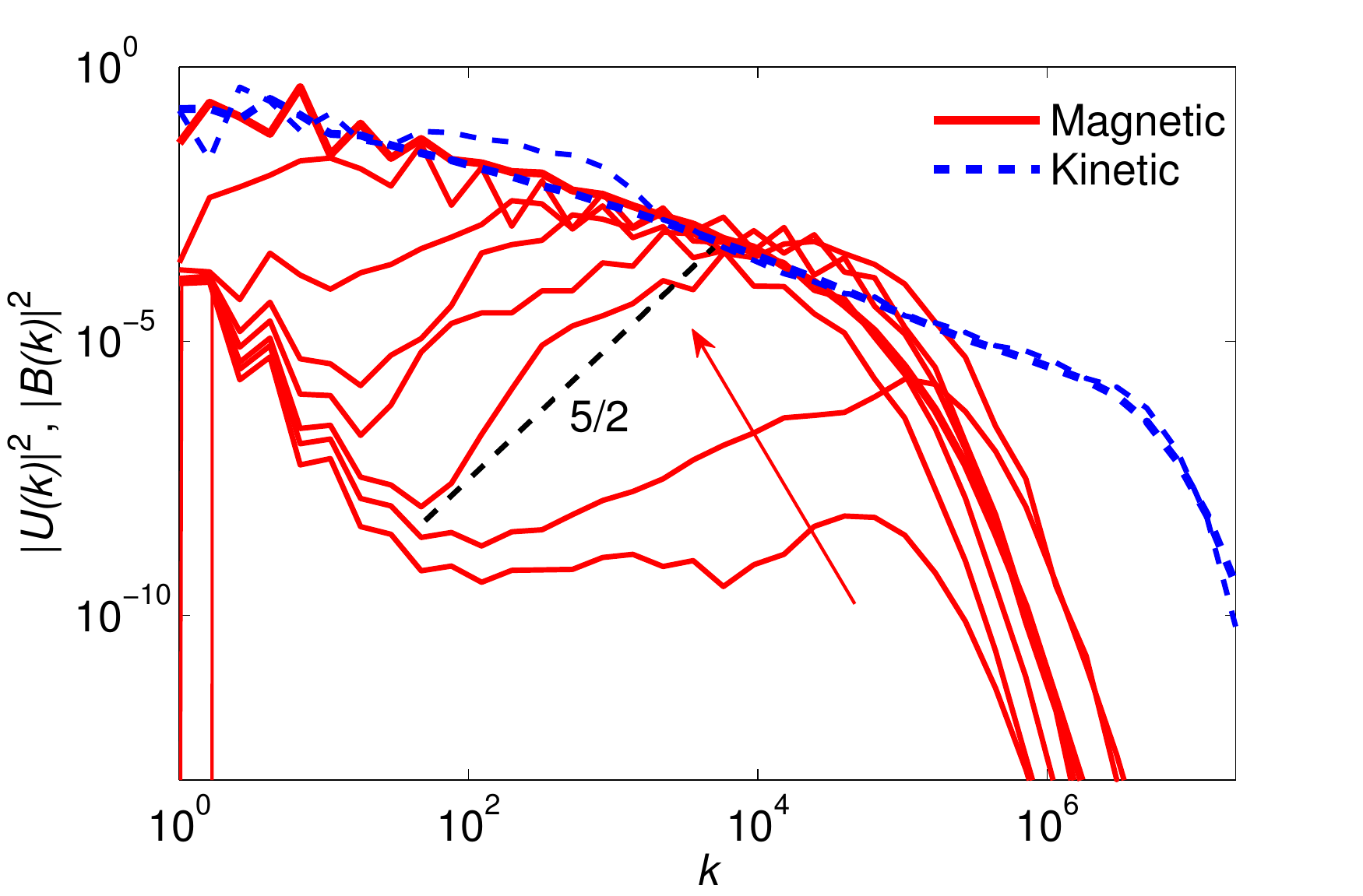}
\caption{For dynamo simulation with $\mathrm{Pm} = 10^{-3}$: Evolution of kinetic energy (dashed blue lines) and magnetic energy (solid red lines) spectra with time.  The seed magnetic field at $t=0$ was at the shells 1 and 2. The magnetic field grows first at large wavenumbers and later at small wavenumbers. In the early phase (kinematic regime),  $|B_n|^2 \sim k^{5/2}$ or 
$E_b(k) \sim k^{3/2}$ which corresponds to the Kazantsev scaling.}
\label{fig:energy_spec_time_lsd}
\end{figure}

In Fig. \ref{fig:energy_spec_time_lsd} we plot the evolution of kinetic and magnetic energy spectra.  For both $\mathrm{Pm}$'s, the kinetic energy does not change appreciably.   The magnetic energy spectrum however grows first at large wavenumbers due to nonlinear transfers, during which an approximate scaling of $|B_n|^2 \sim k^{5/2}$ or  the magnetic energy  $E_b(k) \sim k^{3/2}$  for the low wavenumbers, consistent with the spectrum  proposed by  Kazantsev~\cite{Kazantsev:JETP1968}.   Later, $|B_n|^2$ starts to grow at lower wavenumbers.  These results are consistent with the earlier DNS results~(Brandenburg and Subramanian~\cite{Brandenburg:PR2005}, and references therein).

\begin{figure}[htbp]
\centering
\includegraphics[scale=0.48]{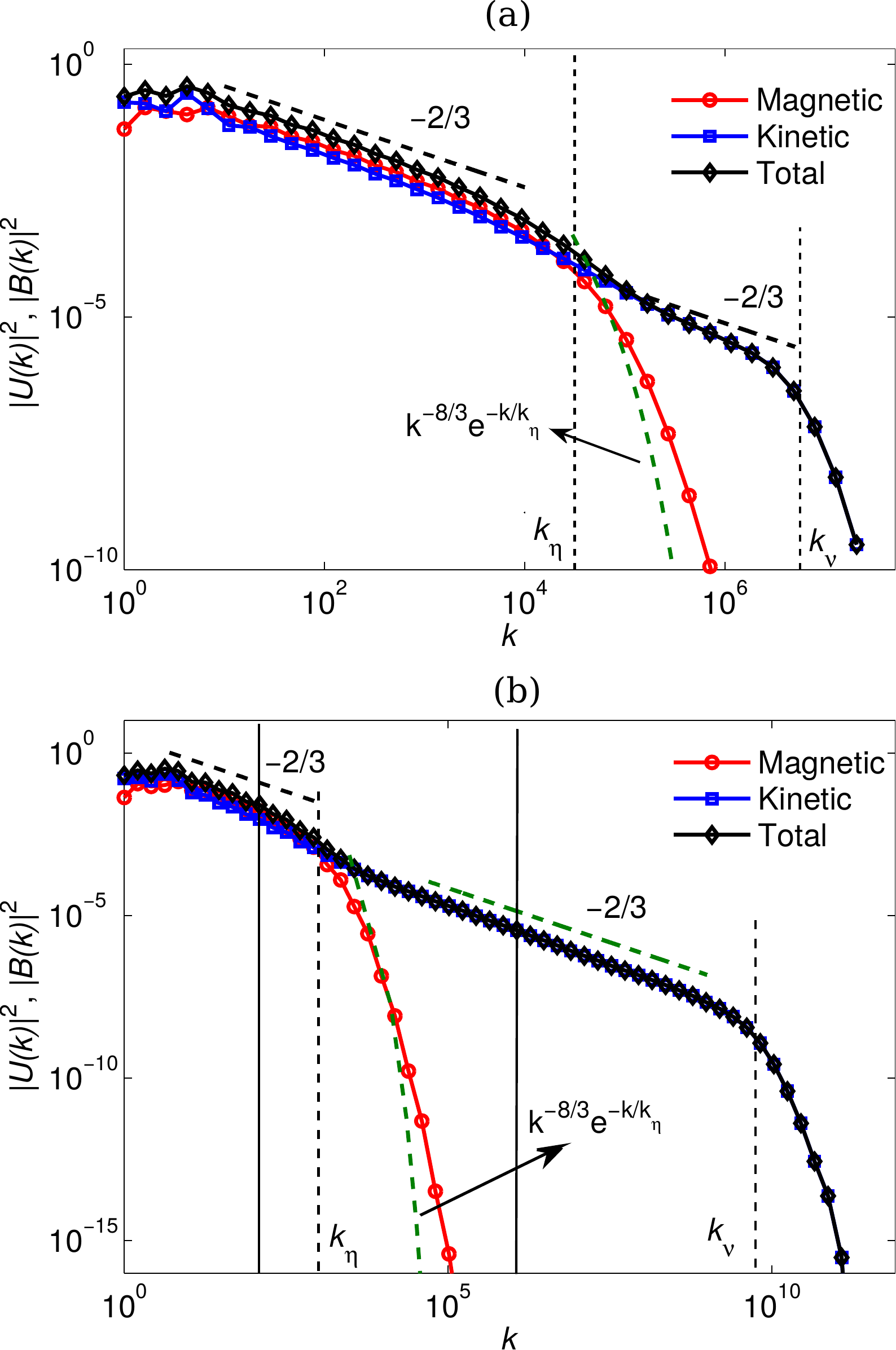}
\caption{For dynamo simulation with (a) $\mathrm{Pm} = 10^{-3}$ and (b) $\mathrm{Pm} = 10^{-9}$: Time-averaged kinetic energy, magnetic energy, and total energy spectra.  The dashed vertical lines represent Kolmogorov wavenumbers $k_\nu$ and $k_\eta$.  For $\mathrm{Pm} = 10^{-9}$, Fig.~\ref{fig:energy_flux_schem_pm_1e-9} exhibits the energy fluxes at $K$'s corresponding to the vertical solid lines.}
\label{fig:energy_spec_small_pm}
\end{figure}

In Fig.~\ref{fig:energy_spec_small_pm}(a,b) we plot the steady-state energy spectra for $\mathrm{Pm}=10^{-3}$ and $10^{-9}$. For $\mathrm{Pm}=10^{-3}$, we observe that $E_u(k)  \sim E_b(k) \sim k^{-5/3}$ curve fits reasonably well in the inertial range, consistent with earlier simulations~~\cite{Stepanov:JT2006,Plunian:NJP2007,Lessinnes:PRE2009}.  For both $\mathrm{Pm}$'s, $E_u(k)   \sim k^{-5/3}$ for $k \lessapprox k_\nu$, but $E_b(k)  \sim k^{-5/3}$ for $k \lessapprox k_\eta$, and $E_b(k)  \sim k^{-11/3} \exp({-k/k_{\eta}})$ for $k_\eta \lessapprox k \lessapprox k_\nu$, as deduced by  Batchelor {\it et al.}~\cite{Batchelor:JFM1959b} and Odier {\em et al.}~\cite{Odier:PRE1998}; the dissipative spectra will be described later in this section.

\begin{figure}[htbp]
\centering
\includegraphics[scale=0.48]{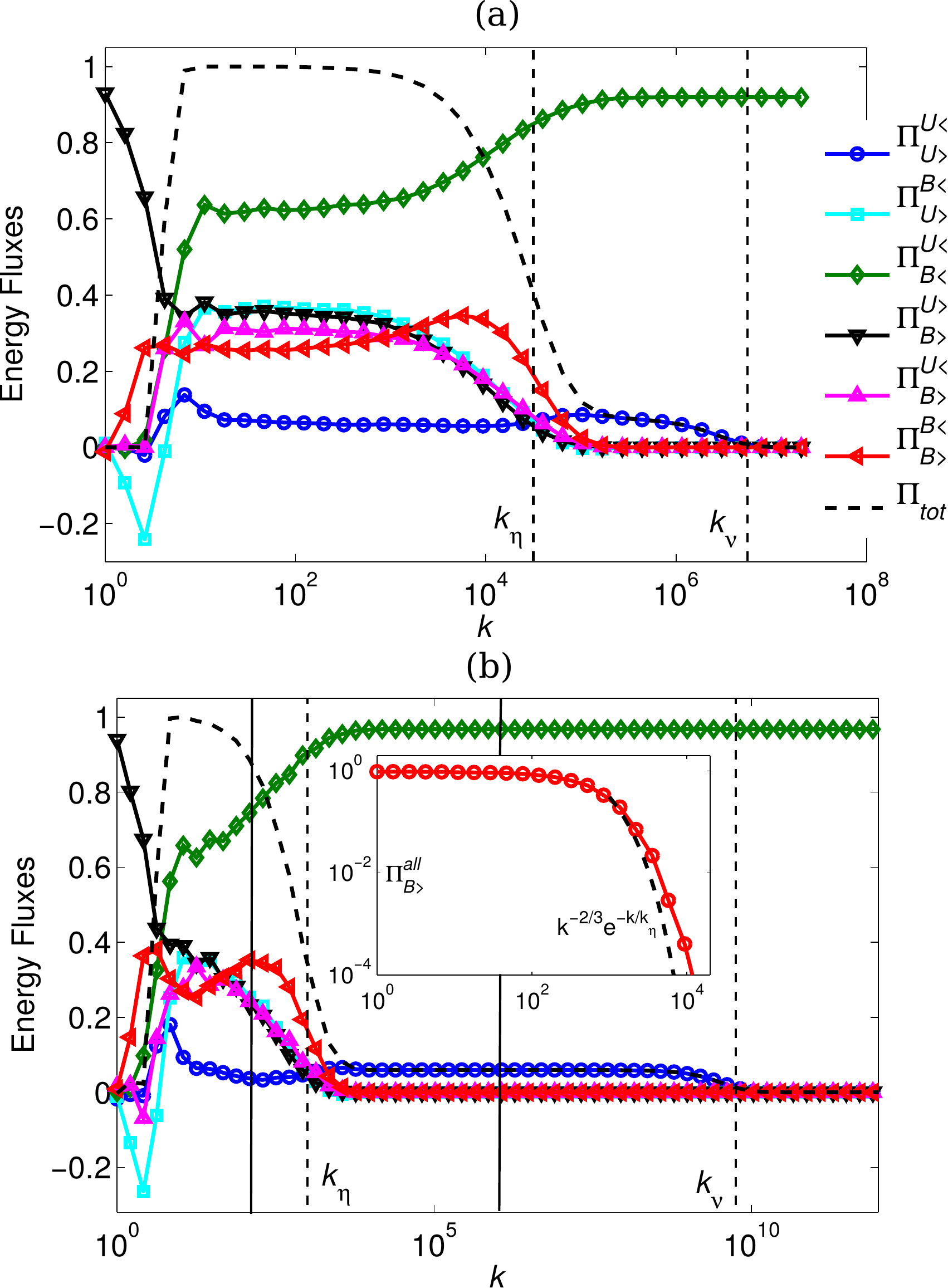}
\caption{For dynamo simulation with (a) $\mathrm{Pm} = 10^{-3}$ and (b) $\mathrm{Pm} = 10^{-9}$: Time-averaged energy fluxes. Subfigure (b) has the same legends as subfigure (a). The inset contains the plot of $\Pi^{all}_{B>} = \Pi^{B<}_{B>} + \Pi^{U<}_{B>} +\Pi^{U>}_{B>}$.   For $\mathrm{Pm} = 10^{-9}$, Fig.~\ref{fig:energy_flux_schem_pm_1e-9} exhibits the energy fluxes at $K$'s corresponding to the vertical solid lines.  }
\label{fig:energy_flux_small_pm}
\end{figure}

We compute various energy fluxes for both the Prandtl numbers.  In the early stages, we observe a rapid energy transfer to the large-$k$ magnetic fields, but in the steady state we observe two broad regimes of energy fluxes as shown in Fig.~\ref{fig:energy_flux_small_pm}(a,b).   The first regime, $ k < k_\eta$, corresponding to the Kolmogorov's $k^{-5/3}$ regime for both $U$ and $B$ field, exhibits similar behaviour as $\mathrm{Pm} =1$ case.  The energy supplied at the small-$k$ of $U$ field gets transferred to the small-$k$ magnetic field, $B \textless$, and the large-$k$ magnetic field, $B \textgreater$. In addition, we have a forward cascade of the magnetic energy that leads to a significant buildup of magnetic energy near the wavenumbers $k \approx k_\eta$ (including a bump). The magnetic energy is dissipated by Joule heating in the second regime, $  k_\eta < k <  k_\nu$.

In Table~\ref{table:flux_small_pm}, we show time-averaged energy fluxes for $\mathrm{Pm} =10^{-3}$ at  $K = 123$ and $4.39\times10^{5}$. For comparison, we also include the shell model results of Lessinnes {\it et al.}~\cite{Lessinnes:PRE2009} for $\mathrm{Pm} =10^{-3}$. We observe that the energy flux $\Pi^{U<}_{U>}$ in our simulation is significantly smaller than that of the Lessinnes {\it et al}. The other major difference is that the energy flux $\Pi^{B<}_{B>}$ in our simulation is positive, whereas it is negative for Lessinnes {\it et al}, i.e., they observed inverse cascade of the magnetic energy. In our shell model simulations, we do not observe any inverse cascade of the magnetic energy, which is consistent with the DNS results of Kumar {\it et al.}~\cite{Kumar:JT2015}. As described earlier, the difference in the $\Pi^{B<}_{B>}$ flux arises due to different form of nonlinearity in the two models. In Table~\ref{table:flux_small_pm_2}, we show energy fluxes for $\mathrm{Pm} =10^{-9}$ at $K =123$ and $1.15\times10^{6}$.  

\begin{figure}[htbp]
\centering
\includegraphics[scale=0.27]{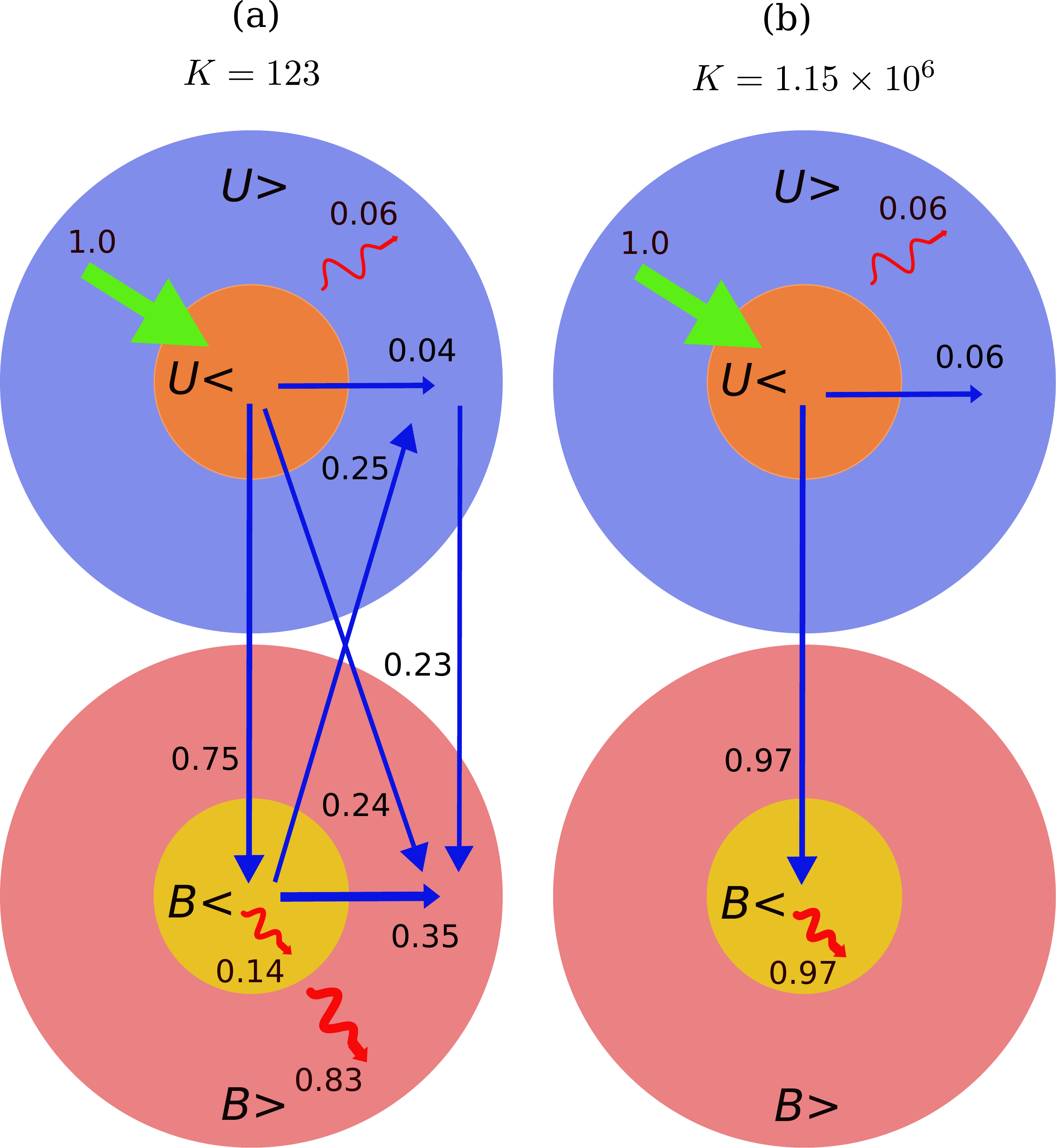}
\caption{For dynamo simulation with $\mathrm{Pm} = 10^{-9}$: Schematic diagrams of the time-averaged energy fluxes and dissipation rates for $K = 123$ and $K = 1.15\times10^{6}$.  The wavy red lines represent viscous and Joule dissipation rates.}
\label{fig:energy_flux_schem_pm_1e-9}
\end{figure}

As shown in Fig.~\ref{fig:energy_flux_schem_pm_1e-9}(b) and Table~\ref{table:flux_small_pm_2}, for $k \gtrapprox k_\eta$,  $\Pi^{U<}_{B<} \approx 97\%$ of the total energy feed; this magnetic energy gets dissipated into Joule heating.   A small fraction of input energy is transferred to small-scale velocity field via  $\Pi^{U<}_{U>}$ ($\approx 6\%$) that gets dissipated via viscous damping.   In summary, most of the input energy is transferred to the magnetic field that gets dissipated by Joule heating.

Now we describe the phenomenology of small-$\mathrm{Pm}$ MHD turbulence. In the inertial range where nonlinear terms of both Eqs.~(\ref{eq:shell_u},\ref{eq:shell_b}) dominate, the energy spectrum follows Kolmogorov-like spectrum~\cite{Verma:PP1999,Verma:PRE2001}.  However, for $k_\eta \lessapprox k \lessapprox k_\nu$,  the kinetic energy continues to have $k^{-5/3}$ energy spectrum, but the magnetic energy spectrum becomes steeper. Odier {\em et al.}~\cite{Odier:PRE1998} deduced that $E_b(k) \sim k^{-11/3}$ using scaling arguments similar to that of Batchelor {\it et al.}~\cite{Batchelor:JFM1959b}. In the regime $k_\eta \lessapprox k \lessapprox k_\nu$,  Odier {\em et al.}~\cite{Odier:PRE1998}  matched $N_n(B,U)$  and the Joule dissipation term of Eq.~(\ref{eq:shell_b}) that yields\begin{equation}
B_{k_\eta} k U_k \sim \eta k^2 B_{k}.
\label{eq:Eb_smallPm_dar}
\end{equation}	
In the language of energy transfers, $\Pi^{B<}_{B>}$ is of the same order as the local Joule dissipation $\eta k^2 |B_k|^2$~\cite{Dar:PD2001}:
\begin{equation}
k  B_{k_\eta} U_k B_k \sim \eta k^2 B_{k}^2.
\end{equation}	
The kinetic energy continues to follow $K^{-5/3}$ spectrum due to the strong $N_n(U,U)$ term.  Using Eq.~(\ref{eq:Eb_smallPm_dar}) and $U_k^2/k \sim \epsilon^{2/3} k^{-5/3}$, Odier {\em et al.}~\cite{Odier:PRE1998} deduced that 
\begin{eqnarray}
E_b(k)  & = & \frac{ B_k^2}{k}  \sim \left( \frac{ B_{k_\eta} }{\eta k} \right)^2 \frac{ U_k^2}{k} 
\nonumber \\
& \sim & \left( \frac{ B_{k_\eta} }{\eta} \right)^2 \epsilon^{2/3} k^{-11/3}.
\end{eqnarray}	
Our numerical data however show steeper spectrum than $k^{-11/3}$ due to the Joule dissipation $\eta k^2 E_b(k)$.  To account for the dissipation, we modify the above spectrum to
\begin{eqnarray}
E_b(k) \sim   \left( \frac{ B_{k_\eta} }{\eta} \right)^2 \epsilon^{2/3} k^{-11/3} \exp(-k/k_\eta),
\label{eq:Eb_smallPm}
\end{eqnarray}
that matches with the numerical data quite well, as shown in Fig.~\ref{fig:energy_spec_small_pm}.  Monchaux {\it et al.}~\cite{Monchaux:PF2009} report  $k^{-11/3}$ and $k^{-17/3}$ spectra at different regimes of $E_b(k)$ of the von K{\'a}rm{\'a}n sodium (VKS) experiment.  They attribute the difference to the advection of the magnetic field either by eddies of the length scales $1/k_\eta$ or $1/k$.   

Since the magnetic energy flux is dissipated by the Joule dissipation, we obtain~\cite{Verma:PF2015}
\begin{equation}
 \frac{d\Pi^{all}_{B>}}{dk} = -2 \eta k^2 E_b(k),
\end{equation}
where $\Pi^{all}_{B>} = \Pi^{B<}_{B>} + \Pi^{U<}_{B>} +\Pi^{U>}_{B>}$ is the total energy flux to the magnetic field at small scales.  Using Eq.~(\ref{eq:Eb_smallPm})  we can argue that
\begin{equation}
\Pi^{all}_{B>}(k)  \sim  2 \eta k^3 E_b(k) \sim k^{-2/3} \exp({-k/k_{\eta}}).
\end{equation}
We observe that $\Pi^{all}_{B>}(k)$ is in qualitative agreement with the above scaling (see inset of Fig.~\ref{fig:energy_flux_small_pm}(b)).  
 
\begin{table}[htbp]
\caption{Time-averaged energy fluxes, viscous dissipation rate, and Joule dissipation rate for $\mathrm{Pm} =10^{-3}$ for our shell model. We also list the results of the shell model  by Lessinnes {\it et al.}~\cite{Lessinnes:PRE2009} for $\mathrm{Pm} =10^{-3}$.} 
\vspace{0.2cm}
\centering
\begin{tabular}{c | c c c}
\hline 
\hline
	 &   \multicolumn{3}{c}{$\mathrm{Pm} = 10^{-3}$} $\rule{0pt}{3ex}$ \\ 
  Flux & Lessinnes shell model & Our shell model & Our shell model \\  
	  & ($K \approx 130$) & ($K = 123$) & ($K = 4.39\times10^{5}$) \\ [1ex]
\hline 
$\Pi^{U<}_{U>}$ & $0.37$ & $0.07$ & $0.07$  \rule{0pt}{3ex} \\

$\Pi^{U<}_{B>}$ & $0.38$ & $0.31$ & $0.00$  \rule{0pt}{4ex} \\

$\Pi^{B<}_{U>}$ & $0.42$ & $0.36$ & $0.00$  \rule{0pt}{4ex} \\

$\Pi^{B<}_{B>}$ & $-0.17$ & $0.26$ & $0.00$  \rule{0pt}{4ex} \\

$\Pi^{U<}_{B<}$ & $0.25$ & $0.62$ & $0.92$  \rule{0pt}{4ex} \\

$\Pi^{U>}_{B>}$ & $0.46$ & $0.35$ & $0.00$ \rule{0pt}{4ex} \\

$\epsilon_\nu(U \textless)$ & $0.00$ & $0.00$ & $0.00$  \rule{0pt}{4ex} \\

$\epsilon_\eta(B \textless)$ & $0.00$ & $0.00$ & $0.92$ \rule{0pt}{3ex} \\

$\epsilon_\nu(U \textgreater)$ & $0.33$ & $0.08$ & $0.07$ \rule{0pt}{3ex} \\

$\epsilon_\eta(B \textgreater)$ & $0.67$ & $0.92$ & $0.00$  \rule{0pt}{3ex} \\
[0.5ex]
\hline
\hline
\end{tabular}
\label{table:flux_small_pm} 
\end{table}

\begin{table}[htbp]
\caption{Time-averaged energy fluxes, viscous dissipation rate, and Joule dissipation rate for $\mathrm{Pm} =10^{-9}$ for our shell model.} 
\vspace{0.2cm}
\centering
\begin{tabular}{c | c c c | c c}
\hline 
\hline
	 &  \multicolumn{2}{c}{$\mathrm{Pm} = 10^{-9}$} $\rule{0pt}{3ex}$ \\  
Flux &  Our shell model & Our shell model \\  
	  & ($K = 123$) & ($K = 1.15\times10^{6}$)  \\ [1ex]
	
\hline 
$\Pi^{U<}_{U>}$ & $0.04$ & $0.06$ \rule{0pt}{3ex} \\

$\Pi^{U<}_{B>}$ & $0.24$ & $0.00$ \rule{0pt}{4ex} \\

$\Pi^{B<}_{U>}$ & $0.25$ & $0.00$ \rule{0pt}{4ex} \\

$\Pi^{B<}_{B>}$ & $0.35$ & $0.00$ \rule{0pt}{4ex} \\

$\Pi^{U<}_{B<}$ & $0.75$ & $0.97$ \rule{0pt}{4ex} \\

$\Pi^{U>}_{B>}$ & $0.23$ & $0.00$ \rule{0pt}{4ex} \\

$\epsilon_\nu(U \textless)$ & $0.00$ & $0.00$ \rule{0pt}{4ex} \\

$\epsilon_\eta(B \textless)$ & $0.14$ & $0.97$ \rule{0pt}{3ex} \\

$\epsilon_\nu(U \textgreater)$ & $0.06$ & $0.06$ \rule{0pt}{3ex} \\

$\epsilon_\eta(B \textgreater)$ & $0.83$ & $0.00$ \rule{0pt}{3ex} \\
[0.5ex]
\hline
\hline
\end{tabular}
\label{table:flux_small_pm_2} 
\end{table}

In the next section we will  simulate dynamos with large Prandtl numbers using our shell model.

\section{Dynamo with large Prandtl numbers}
\label{sec:dyn_large_pm}

Schekochihin {\em et al.}~\cite{Schekochihin:APJ2004}, Stepanov and Plunian~\cite{Stepanov:JT2006}, Buchlin~\cite{Buchlin:AA2011},  Kumar {\it et al.}~\cite{Kumar:EPL2013}, and others showed that the nonlocal energy transfers play a major role in such dynamos during the dynamo growth phase.  Critical nonlocal interactions are missing in our shell model.  Still we use the local shell model described in Sec.~\ref{sec:shell_mod_description} to simulate large-Pm dynamo and explore how well the local shell model captures the dynamo growth and magnetic energy fluxes of large-Pm dynamo.

\begin{figure}[htbp]
\centering
\includegraphics[scale=0.55]{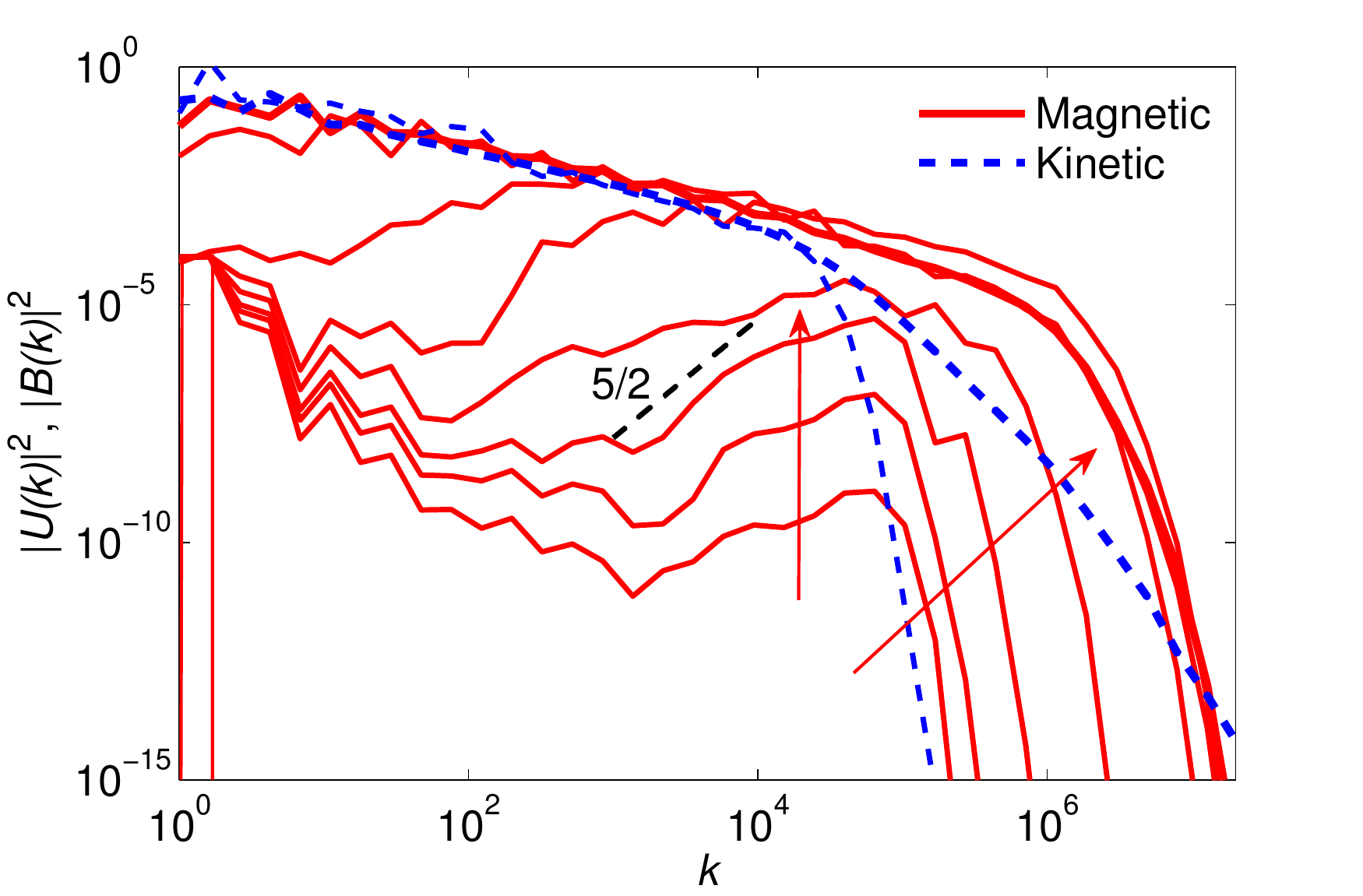}
\caption{For dynamo simulation with $\mathrm{Pm} = 10^{3}$: Evolution of kinetic energy (dashed blue lines) and magnetic energy (solid red lines) spectra with time. $|B_n|^2 \sim k^{5/2}$ corresponds to the Kazantsev prediction at the early stages.}
\label{fig:energy_spec_time_ssd}
\end{figure}

We performed dynamo simulations for $\mathrm{Pm}=10^3$ and $10^9$. Similar to small  $\mathrm{Pm}$ runs, we start our dynamo simulation with the steady-state velocity field of the fluid shell model and a random seed magnetic field.  The magnetic energy first spreads to the larger wavenumber, and then it grows at smaller wavenumber as well.  The evolution of the energy spectrum, shown in Fig.~\ref{fig:energy_spec_time_ssd}, has  strong similarities with the small-$\mathrm{Pm}$ dynamo.  Initially the magnetic energy is quickly transferred from small wavenumbers to larger wavenumbers via local transfers.  After sometime, the magnetic energy grows at large $k$ (small scales) possibly due to weaker magnetic diffusion.  Thus, some features of  the kinematic growth and saturation are captured by the local shell model.  However, nonlocal energy transfers from the small-wavenumber velocity shells to  large-wavenumber magnetic shells are missing in the evolution, which are captured only by nonlocal shell models~\cite{Stepanov:APJ2008}.

\begin{figure}[htbp]
\centering
\includegraphics[scale=0.45]{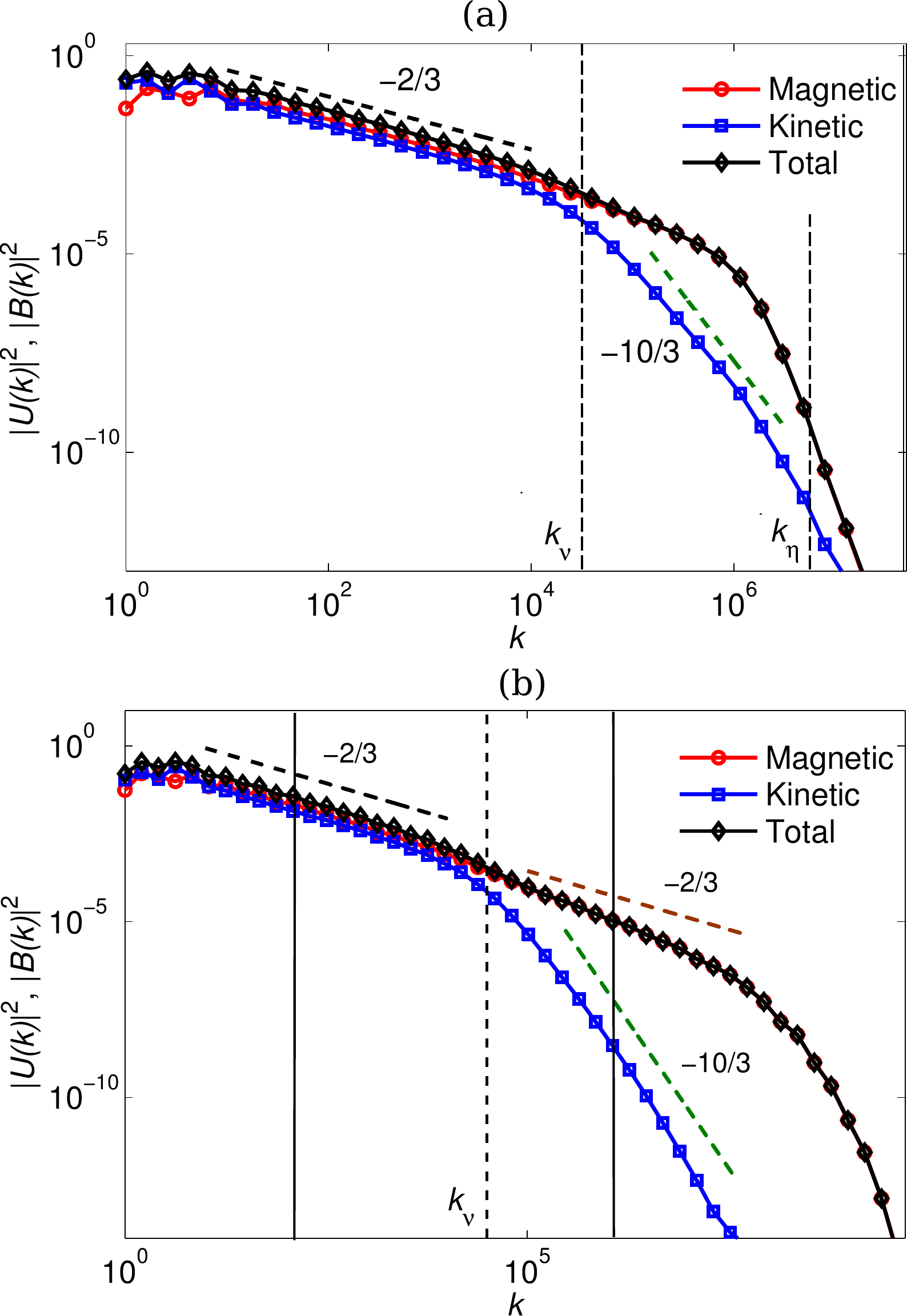}
\caption{For dynamo simulation with (a) $\mathrm{Pm} = 10^{3}$ and (b) $\mathrm{Pm} = 10^{9}$: Time-averaged kinetic energy, magnetic energy, and total energy spectra.  For $\mathrm{Pm} = 10^{9}$, Fig.~\ref{fig:energy_flux_schem_pm_1e9} exhibits the energy fluxes at $K$'s corresponding to the vertical solid lines.}
\label{fig:energy_spec_large_pm}
\end{figure}

We studied the energy spectra for the steady state dynamos with $\mathrm{Pm}=10^3$ and $10^9$.  We illustrate these spectra in Fig.~\ref{fig:energy_spec_large_pm}(a,b).  Note that $k_\nu < k_\eta$ for these runs. The plots of Fig.~\ref{fig:energy_spec_large_pm} reveal that $E_u(k) \sim E_b(k) \sim k^{-5/3}$ in the inertial range with $k \lessapprox k_\nu$. However for $k_\nu \lessapprox k \lessapprox k_\eta$, the magnetic energy spectrum continues to scale as $E_b(k) \sim k^{-5/3}$, but $E_u(k) \sim k^{-13/3}$ due to the following reasons.

For large Pm, the momentum equation can be approximated by
\begin{equation}
\cancel{\frac{\partial {\bf U}}{\partial t} + {\bf U} \cdot {\nabla} {\bf U}}=  -\nabla p + {\bf B} \cdot \nabla {\bf B} + \nu \nabla^2 {\bf U},
\end{equation}
because the viscous term dominates the flow. Here $p$ is the pressure. Performing dimensional analysis, we obtain
\begin{equation}
k  B_k^2 \sim \nu k^2 U_k.
\end{equation}
Using $E_b(k) = B_k^2/k \sim \epsilon^{2/3} k^{-5/3}$ and the above expression, the kinetic energy spectrum is
\begin{equation}
E_u(k) \sim  \frac{U_k^2}{k} \sim \frac{\epsilon^{4/3}}{\nu^2} k^{-13/3}.
\label{eq:Eu_largePm}
\end{equation}
Hence we expect from the shell model that $|U_k|^2 \sim k^{-10/3}$, which is nicely borne out in our simulation results with $|U_k|^2 \sim k^{-3.3 \pm 0.2}$.  The above scaling is very similar to those of Pandey {\it et al.}~\cite{Pandey:PRE2014} derived for  turbulent convection under infinite thermal Prandtl number.

\begin{figure}[htbp]
\centering
\includegraphics[scale=0.45]{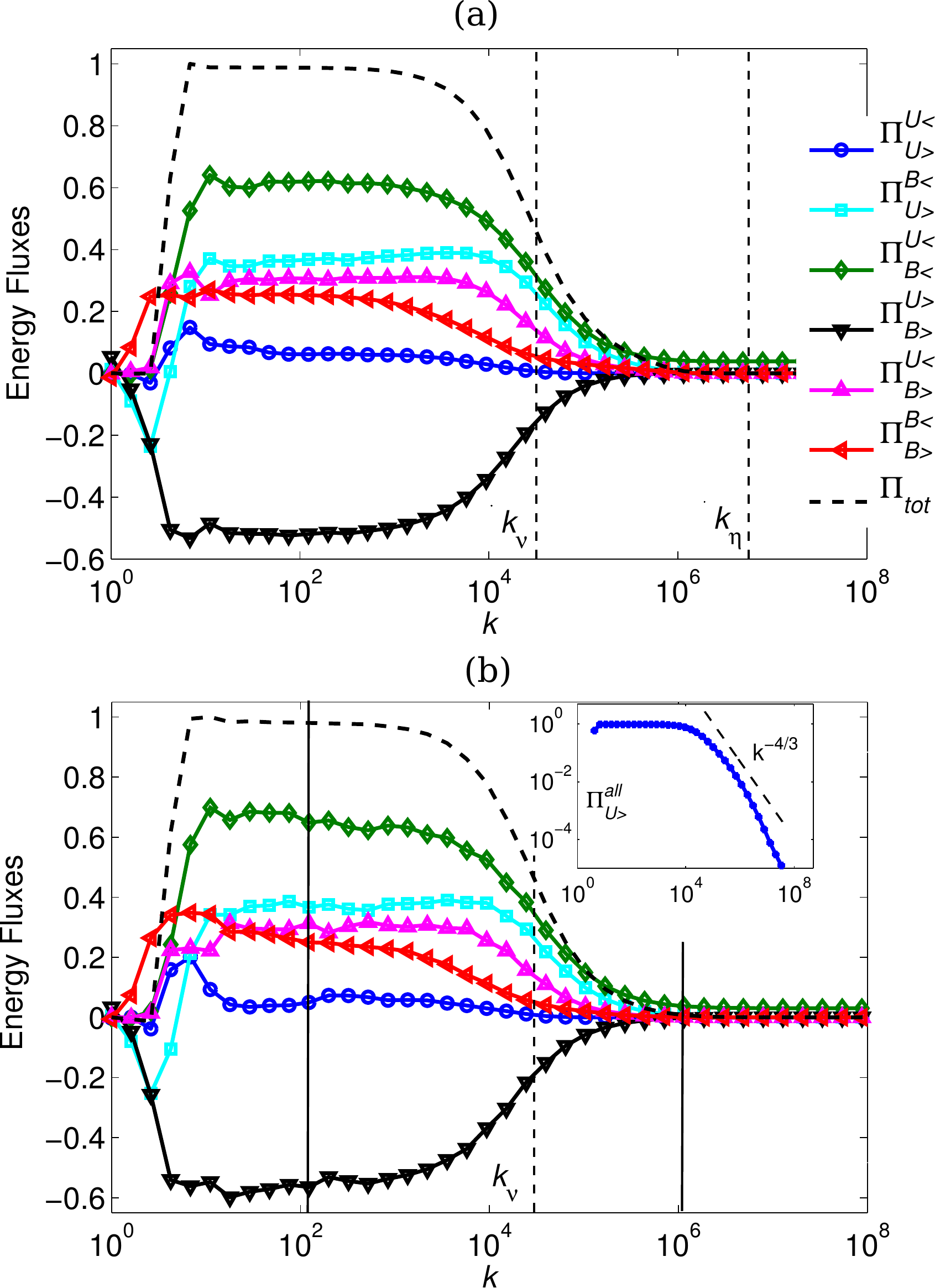}
\caption{For dynamo simulation with (a) $\mathrm{Pm} = 10^{3}$ and (b) $\mathrm{Pm} = 10^{9}$: Time-averaged energy fluxes. Subfigure (b) has the same legends as subfigure (a). The inset contains the plot of $\Pi^{all}_{U>} = \Pi^{U<}_{U>} + \Pi^{B<}_{U>} +\Pi^{B>}_{U>}$.  For $\mathrm{Pm} = 10^{9}$, Fig.~\ref{fig:energy_flux_schem_pm_1e9} exhibits the energy fluxes at $K$'s corresponding to the vertical solid lines. }
\label{fig:energy_flux_large_pm}
\end{figure}

We compute various energy fluxes for both the runs during the steady state by averaging them over a long time. The simulation results are exhibited in Fig.~\ref{fig:energy_flux_large_pm}.  The energy fluxes for both the Prandtl numbers appear to be similar.  The  fluxes take constant values in the inertial range, after which all the fluxes  tend to zero. The net energy flux to the velocity field, $\Pi^{all}_{U>} = \Pi^{U<}_{U>}+\Pi^{B<}_{U>}+\Pi^{B>}_{U>}$, is dissipated by the viscous dissipation, hence 
\begin{equation}
 \frac{d\Pi^{all}_{U>}}{dk} = -2 \nu k^2 E_u(k).
\end{equation}
 Using Eq.~(\ref{eq:Eu_largePm}) we can argue that  
 \begin{equation}
\Pi^{all}_{U>}  \sim 2 \nu k^3 E_u(k) \sim k^{-4/3}.
\end{equation}
which is qualitatively borne out by $\Pi^{all}_{U>}$ as well dominant fluxes as shown in Fig.~\ref{fig:energy_flux_large_pm}.
  
\begin{figure}[htbp]
\centering
\includegraphics[scale=0.27]{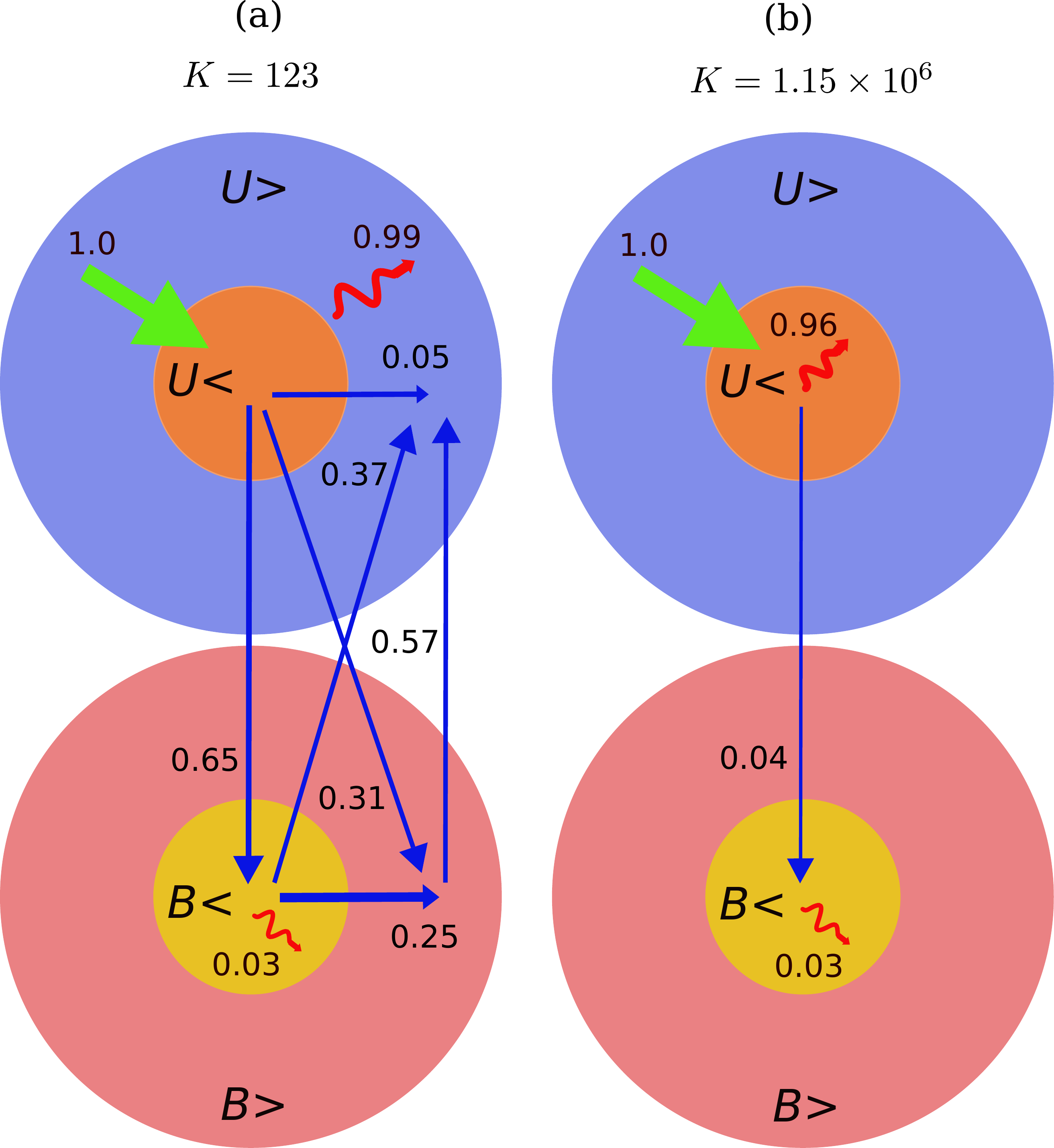}
\caption{For dynamo simulation with $\mathrm{Pm} = 10^{9}$: Schematic diagram of the time-averaged energy fluxes and dissipation rates at $K = 123$ and $K = 1.15\times10^{6}$.}
\label{fig:energy_flux_schem_pm_1e9}
\end{figure}

In Fig.~\ref{fig:energy_flux_schem_pm_1e9} we display the constant values of the energy fluxes for $\mathrm{Pm} = 10^{9}$ at $K=123$ and $K=1.15\times10^6$.  The kinetic energy at small wavenumber is transferred  to $B \textless$ and $B \textgreater$, which in turn gets transferred to $U \textgreater$ (large wavenumber). This energy is dissipated via viscous damping.  Interestingly, the direct cascade of kinetic energy to small scales ($\Pi^{U<}_{U>}$) is very small; the growth of the kinetic energy at small scales occur via kinetic-to-magnetic and then back to kinetic energy transfers.  Almost all the energy fed at small wavenumbers of the velocity field is dissipated at large wavenumbers of the velocity field by viscous dissipation.  The Joule dissipation is insignificant for large-$\mathrm{Pm}$ dynamo. In contrast, maximal dissipation in small-$\mathrm{Pm}$ dynamos occurs via Joule heating. 

In Table~\ref{table:flux_large_pm}, we show time-averaged energy fluxes for $\mathrm{Pm} =10^{3}$ at $K = 123$, and for $\mathrm{Pm} =10^{9}$ at $K =123$ and $1.15\times10^{6}$. For comparison, we also list DNS results for $\mathrm{Pm} =20$ by Kumar {\it et al.}~\cite{Kumar:EPL2013}. The energy fluxes in our shell model simulation are in qualitative agreement with the DNS results of Kumar {\it et al.}, but with certain major differences.   In our shell model, $\Pi^{U<}_{B<}$ is the most dominant energy flux for the magnetic energy growth, whereas in DNS the magnetic energy grows primarily due to $\Pi^{U<}_{B>}$. The deviation from Kumar {\it et al.} is due to the absence of nonlocal interactions in our shell model. A nonlocal shell model should be able to capture the aforementioned nonlocal energy transfers.

\begin{figure}[htbp]
\centering
\includegraphics[scale=0.5]{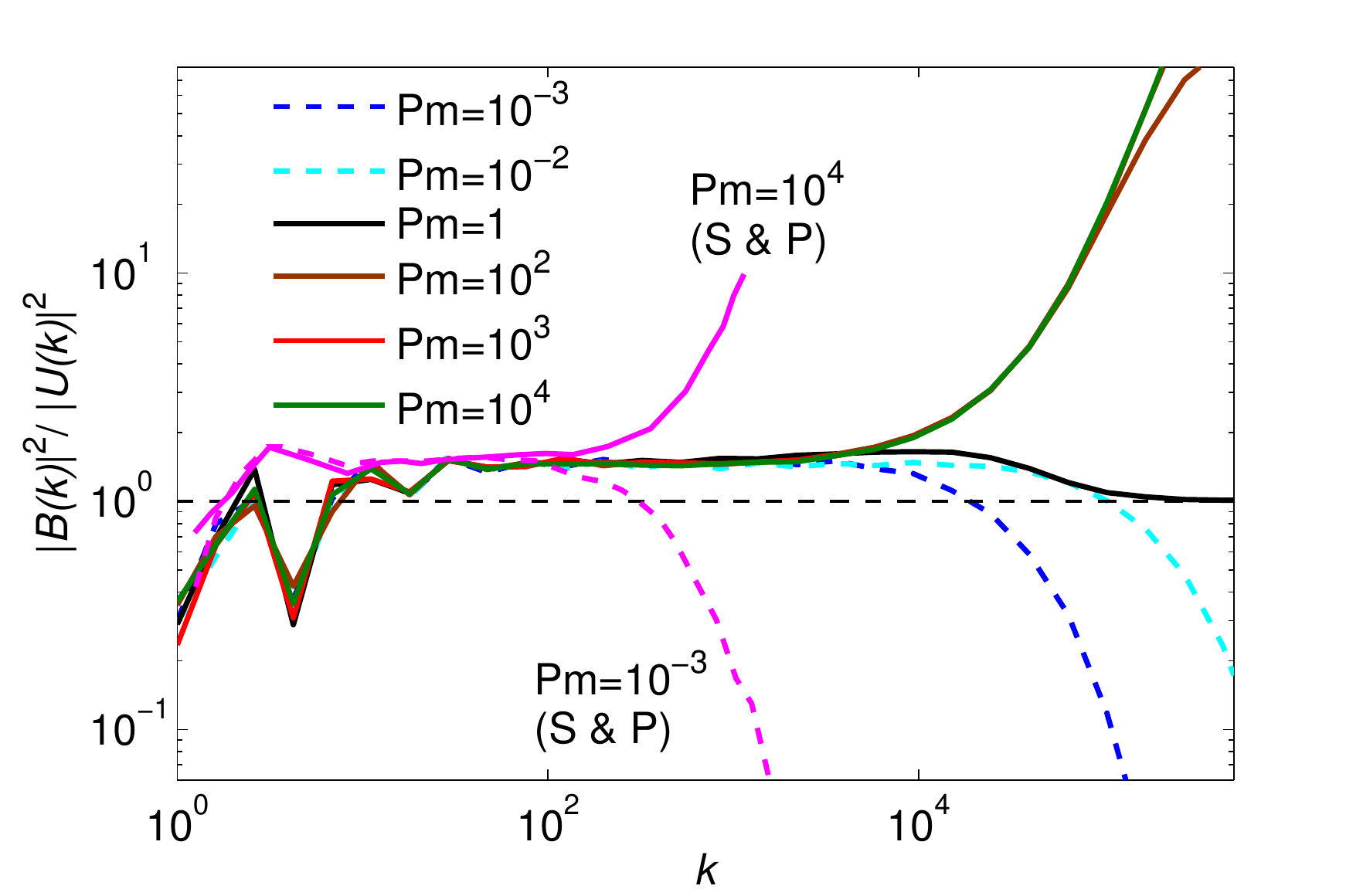}
\caption{For $\mathrm{Pm} \in [10^{-3}, 10^4]$: The ratio of magnetic energy ($|B(k)|^2$) and kinetic energy ($|U(k)|^2$) with wavenumber. The lines with magenta colour represent the ratio computed for $\mathrm{Pm} =10^{-3}$ and $10^{4}$ by Stepanov and Plunian (S $\&$ P)~\cite{Stepanov:APJ2008}.}
\label{fig:ratio_euk_ebk}
\end{figure}

We summarize our spectra results by plotting $E_b(k)/E_u(k)$ for  various Prandtl numbers (see Fig.~\ref{fig:ratio_euk_ebk}).  We observe that the $E_b(k)/E_u(k) \approx 1.5$ in the regime where both $E_u(k) \sim E_b(k) \sim k^{-5/3}$.  For large $\mathrm{Pm}$, $E_b(k)/E_u(k)$ grows monotonically with $k$, but it decreases with $k$ for small $\mathrm{Pm}$ due to the nature of energy spectra described in the last two sections. Our results are in general agreement with the results of Stepanov and Plunian~\cite{Stepanov:APJ2008}. Note that the inertial range in our simulation is broader because our viscosity and the magnetic diffusivity are smaller than those of Stepanov and Plunian~\cite{Stepanov:APJ2008}. 

\begin{table*}[htbp]
\caption{Time-averaged energy fluxes, viscous dissipation rate, and Joule dissipation rate for $\mathrm{Pm} =10^3$ and $10^9$ for our shell model.  We also list the DNS results of Kumar {\it et al.}~\cite{Kumar:EPL2013} for $\mathrm{Pm} =20$ .} 
\vspace{0.2cm}
\centering
\begin{tabular}{c | c | c | c c}
\hline 
\hline 
	 & $\mathrm{Pm} = 20$ & $\mathrm{Pm} = 10^{3}$ & \multicolumn{2}{c}{$\mathrm{Pm} = 10^{9}$} $\rule{0pt}{3ex}$ \\	 
	Flux & DNS (Kumar)  & Shell model & Shell model & Shell model \\ 
		 & ($K \approx 22$)	  	  & ($K = 123$)    & ($K = 123$)        & ($K = 1.15\times10^{6}$) \\[1ex]
	
\hline 
$\Pi^{U<}_{U>}$ & $0.01$ & $0.06$ & $0.05$ & $0.00$ \rule{0pt}{3ex} \\

$\Pi^{U<}_{B>}$ & $0.63$ & $0.31$ & $0.31$ & $0.00$ \rule{0pt}{4ex} \\ 

$\Pi^{B<}_{U>}$ & $0.00$ & $0.37$ & $0.37$ & $0.00$ \rule{0pt}{4ex} \\

$\Pi^{B<}_{B>}$ & $0.36$ & $0.25$ & $0.25$ & $0.00$ \rule{0pt}{4ex} \\

$\Pi^{U<}_{B<}$ & $0.57$ & $0.62$ & $0.65$ & $0.04$ \rule{0pt}{4ex} \\

$\Pi^{U>}_{B>}$ & $-0.11$ & $-0.52$ & $-0.57$ & $0.00$ \rule{0pt}{3ex} \\

$\epsilon_\nu(U \textless)$ & $-$ & $0.01$ & $0.00$ & $0.96$ \rule{0pt}{4ex} \\

$\epsilon_\eta(B \textless)$ & $-$ & $0.00$ & $0.03$ & $0.03$ \rule{0pt}{3ex} \\

$\epsilon_\nu(U \textgreater)$ & $-$ & $0.95$ & $0.99$ & $0.00$ \rule{0pt}{3ex} \\

$\epsilon_\eta(B \textgreater)$ & $-$ & $0.04$ & $0.00$ & $0.00$ \rule{0pt}{3ex} \\
[0.5ex]
\hline
\hline
\end{tabular}
\label{table:flux_large_pm} 
\end{table*}

\section{Discussion and Conclusions}
\label{sec:conclu}

In this paper we presented an MHD shell model that is suitable for computing energy fluxes.  We derived formulas for various fluxes of dynamo.  The fluxes of our shell model are in good agreement with those computed using  direct numerical simulations for the magnetic Prandtl number $\mathrm{Pm}=1$~\cite{Debliquy:PP2005}; thus we validate the shell model for flux computations.  We employ this model to study energy spectra  and fluxes for very small and very large magnetic Prandtl numbers (from $\mathrm{Pm} = 10^{-9}$ to $10^9$), which are inaccessible to direct numerical simulations. Our major findings are as follows:
\begin{enumerate}
\item For small $\mathrm{Pm}$, the kinetic energy spectrum $E_u(k) \sim k^{-5/3}$, but the magnetic energy spectrum $E_b(k) \sim k^{-5/3}$ for $k < k_\eta = (\epsilon/\eta^3)^{1/4}$, and $E_b(k) \sim k^{-11/3}\exp(-k/k_\eta)$ for $k_\eta < k < k_\nu$, where $k_\nu = (\epsilon/\nu^3)^{1/4}$, similar to Odier {\em et al.}~\cite{Odier:PRE1998}.  

\item For small $\mathrm{Pm}$, small-$k$ velocity field feeds small-$k$ magnetic field that cascades to large-$k$ magnetic field where it gets dissipated. Here, most of the dissipation is through Joule heating.

\item The large $\mathrm{Pm}$ dynamo exhibits  $E_b(k) \sim k^{-5/3}$, but $E_u(k) \sim k^{-5/3}$ for $k < k_\nu$ and $E_u(k) \sim k^{-13/3}$ for $k_\nu < k < k_\eta$.

\item For large $\mathrm{Pm}$, small-$k$ velocity field feeds the small-$k$ magnetic field, which in turn cascades to large-$k$ velocity and magnetic field.  The large-$k$ magnetic field transfers energy to the large-$k$ velocity field.  As a result, most of the energy fed to the velocity field by forcing at small-$k$ returns to large-$k$ velocity field where it gets dissipated via viscous force.  
\end{enumerate}

In conclusion, we construct an MHD shell model which we use to compute the energy spectra and fluxes for extreme  $\mathrm{Pm}$'s.  The numerical simulation of this model provides valuable insights into small-$\mathrm{Pm}$  and large-$\mathrm{Pm}$ dynamos. Extensions of our shell model with nonlocal and realistic  helical effects would be very useful for more realistic dynamos.
   

\section*{Acknowledgements}

We thank Abhishek Kumar, Daniele Carati, Thomas Lessinnes, Rodion Stepanov, Franck Plunian, Stephan Fauve, and Sagar Chakraborty for discussions at various stages, and A. G. Chatterjee for the help with post-processing scripts. This work was supported by a research grant SERB/F/3279/2013-14 from Science and Engineering Research Board, India.




\pagebreak

\appendices

\section{Energy transfer and fluxes in fluid shell model}
\label{appendix:fluid}
Our objective is to derive an expression for the energy fluxes in MHD turbulence for which we need to derive a formula for the shell-to-shell energy transfer. For simplicity, first we derive these formulas for the fluid shell model, which is Eq.~(\ref{eq:shell_u}) with $B=0$.  

As discussed in Sec.~\ref{sec:shell_mod_description}, the kinetic energy of this system is conserved when $\nu=0$ and $F_n=0$.  However, we  get further insights into the energy transfers when we focus on a single unit of energy transfers, which is a set of three consecutive shells $(n-1,n,n+1)$.  Note that this unit is analogous to a {\em triad interaction} (${\bf k,p,q}$) with ${\bf k=p+q}$ in Fourier space representation of  the Navier Stokes equation.  The energy equations for the aforementioned three shells are
\begin{eqnarray}
\frac{d|U_{n-1}|^{2}/2}{dt} & = & -a_1 k_{n-1} \Im(U_{n-1}U_{n}U_{n+1})  \nonumber \\
& = & L^{UU}(n-1|n,n+1),  \label{eq:vel_n-1} \\
\frac{d|U_{n}|^{2}/2}{dt} & = & -a_{2} k_{n-1} \Im(U_{n-1}U_{n}U_{n+1}) \nonumber \\ 
 & = & L^{UU}(n|n-1,n+1),  \label{eq:vel_n} \\
\frac{d|U_{n+1}|^{2}/2}{dt} & = & -a_{3} k_{n-1} \Im(U_{n-1}U_{n}U_{n+1}) \nonumber \\
 & = & L^{UU}(n+1|n-1,n),  \label{eq:vel_n+1}
\end{eqnarray}
where $\Im$ denotes the imaginary part of the argument.   Interestingly, the total kinetic energy is conserved in this unit interaction when we impose $a_1+a_2+a_3=0$ [Eq.~(\ref{eq:sum_a_i_0})]. This  is the {\em detailed energy conservation} in a triadic unit of shells.  When all the triads are included, then the total energy would also be conserved. 

\begin{figure}[htbp]
\centering
\includegraphics[scale=2.0]{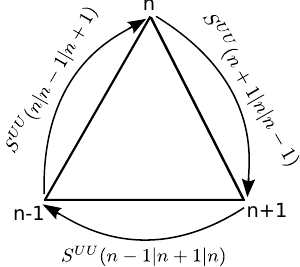}
\caption{Schematic diagram for energy transfers in a triad.}
\label{fig:triad_en_tr}
\end{figure}

Each of the shells $(n-1,n,n+1)$ receives energy from the other two.  Motivated by the {\em mode-to-mode} energy transfer formulas derived by Dar {\it et al.}~\cite{Dar:PD2001} and Verma~\cite{Verma:PR2004}, we set out to explore whether we can derive a formula for the energy transfer to a shell from the other shells.  Let us postulate that $S^{UU}(n|m|p)$ is the energy transfer rate from the shell $m$ to the shell $n$ with the shell $p$ acting as a mediator. Hence,
\begin{eqnarray}
S^{UU}(n+1|n|n-1) + S^{UU}(n+1|n-1|n) =L^{UU}(n+1|n-1,n), \label{eq:S-n+1-fluid} \\
S^{UU}(n|n+1|n-1)+ S^{UU}(n|n-1|n+1) = L^{UU}(n|n-1,n+1), \label{eq:S-n-fluid}\\
S^{UU}(n-1|n+1|n)+ S^{UU}(n-1|n|n+1) = L^{UU}(n-1|n,n+1).\label{eq:S-n-1-fluid}
\end{eqnarray}
Also, the energy received by shell $n$ from the shell $m$ is equal and opposite to the energy received by shell $m$ from the shell $n$.  This condition provides three additional relations for the $S^{UU}$'s, i.e.,
\begin{eqnarray}
S^{UU}(n+1|n|n-1) & = & -S^{UU}(n|n+1|n-1)  \label{eq:1}\\
S^{UU}(n|n-1|n+1) & = & -S^{UU}(n-1|n|n+1) \label{eq:2}\\
S^{UU}(n-1|n+1|n) & = & -S^{UU}(n+1|n-1|n).  \label{eq:3}
\end{eqnarray}
The desired energy transfer formulae $S$'s are linear functions of $U_{n-1}$, $U_n$, and $U_{n+1}$ as well as one of the wavenumber, say $k_{n-1}$. Hence, we choose the following form for $S$'s:
\begin{eqnarray}
S^{UU}(n+1|n|n-1) & = & \alpha_1 A_n, \\
S^{UU}(n|n-1|n+1) & = &  \alpha_2 A_n, \\
S^{UU}(n-1|n+1|n) & = &  \alpha_3 A_n,
\end{eqnarray}
where
\begin{equation}
A_n = k_{n-1} \Im(U_{n-1}U_{n}U_{n+1}).
\end{equation}
These transfers are depicted in Fig.~\ref{fig:triad_en_tr}.  Using Eqs.~(\ref{eq:S-n+1-fluid}-\ref{eq:3}) and the definitions of $S$'s, we obtain
\begin{eqnarray}
\alpha_1 - \alpha_3 & = & -a_3, \\
\alpha_2-\alpha_1 & = &  -a_2 
\end{eqnarray}

Hence the solutions of $S^{UU}$'s are
\begin{eqnarray}
S^{UU}(n+1|n|n-1) & = & \alpha_1 A_n, \label{eq:S-n+1-fluid-soln}\\
S^{UU}(n|n-1|n+1) & = & (\alpha_1-a_2) A_n,  \label{eq:S-n-fluid-soln}\\
S^{UU}(n-1|n+1|n) & = & (\alpha_1+a_3) A_n. \label{eq:S-n-1-fluid-soln}
\end{eqnarray}

For our computations we take $\alpha_1 = \lambda = (\sqrt{5}+1)/2$ after which $\alpha_2$ and $\alpha_3$ are automatically determined. 

The above solution however is not unique; one can add a {\em circulating transfer} that traverses $(n-1) \rightarrow n \rightarrow (n+1) \rightarrow (n-1)$. Following the same arguments as Dar {\it et al.}~\cite{Dar:PD2001} and Verma~\cite{Verma:PR2004}, we can show that the circulating transfer does not affect the value of energy flux, which is defined below.   A shell model contains a large number of shells.  However, all the energy transfers can be split into the unit interactions discussed above.  

The energy flux in fluid turbulence is defined as the energy leaving a wavenumber sphere of radius $K$.  It is easy to define the energy flux $\Pi^{U<}_{U>}(K)$ for a shell model as the energy transfers from all the shells within a sphere of radius $K$ to the shells outside the sphere, which is
\begin{eqnarray}
\Pi_{U>}^{U<}(K) & = &   \sum_{m\le K}  \sum_{n>K} \sum_{p}S^{UU}(n|m|p).
\end{eqnarray}

Note that the giver shell is the superscript of $\Pi$ and the receiver shell is the subscript.  Since the interactions in such a shell model is local, that is, only three neighbouring shells interact, the above energy  flux gets contributions from shells $K-1, K, K+1, K+2$ as given below:
\begin{eqnarray}
\Pi_{U>}^{U<}(K) & = & \sum_{n>K}\sum_{m\le K}\sum_{p}S^{UU}(n|m|p)\nonumber \\
& = & S^{UU}(K+1|K|K-1) +S^{UU}(K+1|K|K+2) \nonumber  \\
&  & +S^{UU}(K+1|K-1|K)  +S^{UU}(K+2|K|K+1) \nonumber \\
& = & \alpha_1 k_{K-1} \Im(U_{K+1}U_{K}U_{K-1})  + \alpha_2 k_{K-1} \Im(U_{K}U_{K-1}U_{K+1}) \nonumber \\
& & -\alpha_3 k_{K-1} \Im(U_{K+1}U_{K-1}U_{K})   -\alpha_3 k_{K-1} \Im(U_{K+1}U_{K-1}U_{K})
\end{eqnarray}

In the next appendix, we will extend this formalism to MHD turbulence.

\section{Energy transfer and fluxes in MHD shell model}
\label{appendix:MHD}
\begin{figure}[htbp]
\centering
\includegraphics[scale=0.70]{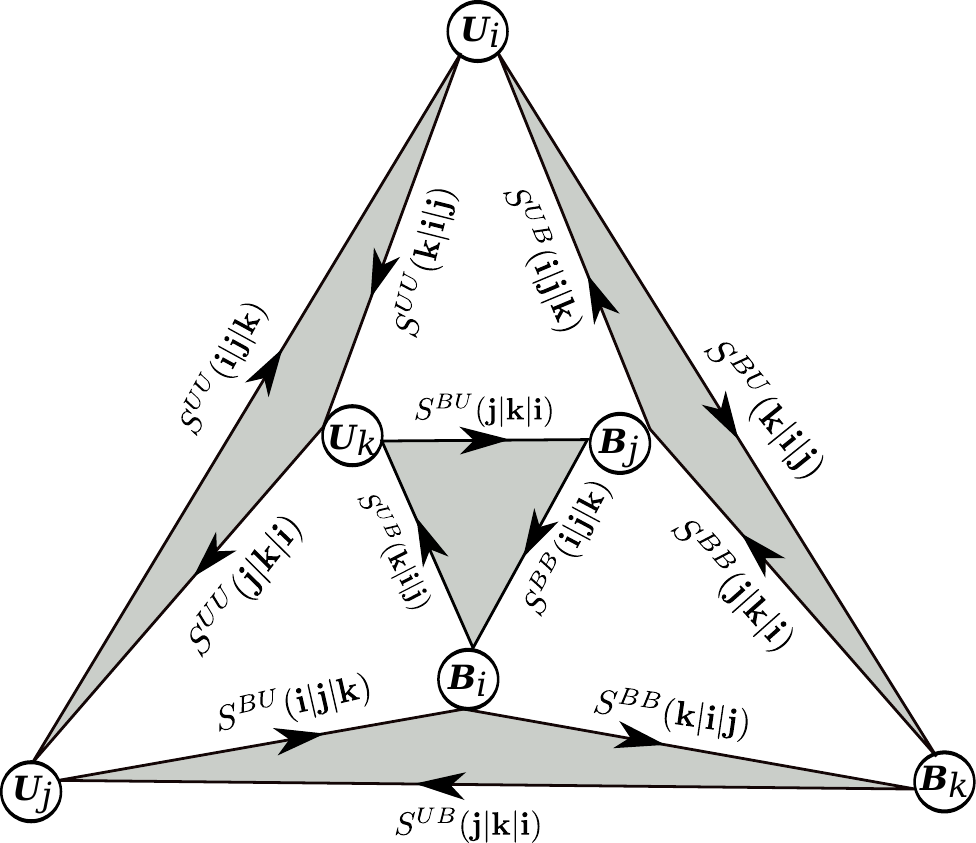}
\caption{Triadic interactions among velocity and magnetic shells. Reprinted from Plunian {\it et al.}~\cite{Plunian:PR2013} copyright (2012), with permission from Elsevier.}
\label{fig:triad_int}
\end{figure}

The derivation for the energy transfers in MHD turbulence is very similar to that for fluid turbulence described in Appendix~\ref{appendix:fluid}.  It  gets more complex due to further interactions between the velocity and magnetic fields.  In this appendix, we will state the final results after  a sketch of the proof.  The derivation follows a similar structure as followed by Dar {\it et al.}~\cite{Dar:PD2001} and Verma~\cite{Verma:PR2004}. In Fig.~\ref{fig:triad_int}, we exhibit a schematic diagram for triadic interactions among velocity and magnetic modes (taken from Plunian {\it et al.}~\cite{Plunian:PR2013}).

To derive the shell-to-shell energy transfers in MHD turbulence we limit ourselves to a unit of energy transfers, which is a combination of three consecutive shells $(n-1, n, n+1)$.  We start with Eqs.~(\ref{eq:shell_u}) and (\ref{eq:shell_b})  and derive the energy equations for the shell variables.  The energy equations for the $n$-th shell are
\begin{eqnarray}
\frac{d|U_{n}|^{2}/2}{dt} & = & -a_{2} k_{n-1} \Im(U_{n-1}U_{n}U_{n+1})  -2b_{2}k_{n-1}\Im(U_{n}B_{n-1}B_{n+1}) \nonumber \\
 & = & L^{UU}(n|n-1,n+1)   +L^{UB}(n|n-1,n+1), \label{eq:mhd_u_energy} \\
 \frac{d|B_{n}|^{2}/2}{dt} & = & b_3 k_{n-1}\Im(B_{n-1}B_{n}U_{n+1})  + b_1 k_{n-1}\Im(U_{n-1}B_{n}B_{n+1})  \nonumber \\
& & + d_3 k_{n-1}\Im(B_{n-1}B_{n}U_{n+1})  - d_2 k_{n-1}\Im(U_{n-1}B_{n}B_{n+1})  \nonumber \\
& = & L^{BU}(n|n-1,n+1)   +L^{BB}(n|n-1,n+1).  \label{eq:mhd_b_energy}
\end{eqnarray}
Here $L^{UU}, L^{BB}, L^{UB}, L^{BU}$ represent the energy transfers from velocity-to-velocity ($U2U$), magnetic-to-magnetic ($B2B$),  magnetic-to-velocity ($B2U$), and velocity to magnetic ($U2B$) respectively. 

The $L^{UU}$ terms follow same dynamics as that for the fluid turbulence, and hence the formulas derived in Appendix~\ref{appendix:fluid} is applicable for the $U2U$ transfers.  It is easy to show that the $B2B$ transfer terms arising due to $N_n[U,B]$ has similar properties as $N_n[U,U]$, that is, it conserves $\sum_n |B_n|^2/2$. Following similar lines as the derivation for fluid shell model, we derive the shell-to-shell transfers for $B2B$ transfers as:
\begin{eqnarray}
S^{BB}(n+1|n|n-1) & = & d_{2}  P^{BB}_{U}(n+1|n|n-1), \label{eq:Sbb-n-mhd-soln}\\
S^{BB}(n|n-1|n+1) & = & d_{3}  P^{BB}_{U}(n|n-1|n+1), \label{eq:Sbb-n-1-mhd-soln}\\
S^{BB}(n-1|n+1|n) & = & (-d_{1}) P^{BB}_{U}(n-1|n+1|n), \label{eq:Sbb-n-2-mhd-soln}
\end{eqnarray}
where 
 \begin{eqnarray}
P^{YX}_{Z}(n|m|p) & = & k_{\min(n,m,p)} \Im (Y_n X_m Z_p),
\label{eq:P}  
\end{eqnarray}
with  the giver shell $X=B$,  the receiver shell $Y=B$, and the mediator shell $Z=U$.  Note that the velocity field acts as a mediator in $B2B$ energy transfers, and it has similar role as the velocity field ${\bf U}$ in the nonlinear interactions of ${\bf U} \cdot \nabla {\bf B}$ in the MHD equations.  

Using the $B2B$ shell-to-shell formula we can derive the energy flux $\Pi^{B<}_{B>}(K)$, the magnetic energy transfers from all the shells within the sphere of radius $K$ to the shells outside the sphere:
\begin{eqnarray}
\Pi_{B>}^{B<}(K) & = &   \sum_{m\le K}  \sum_{n>K}  \sum_{p}S^{BB}(n|m|p).
\label{eq:PiBl_Bg}  
\end{eqnarray}

Now we work on the $U2B$ and $B2U$ energy transfers that occur via nonlinear terms $N_n[B,U]$ and $N_n[B,B]$ of the shell model.  The corresponding terms in the energy Eqs.~(\ref{eq:mhd_u_energy}) and (\ref{eq:mhd_b_energy}) are $L^{UB}$ and $L^{BU}$ respectively.  It is easy to verify that 
\bea
&&L^{BU}(n|n-1,n+1) + L^{UB}(n|n-1,n+1) \nonumber \\
& + &  L^{BU}(n-1|n,n+1) + L^{UB}(n-1|n,n+1) \nonumber \\
&+ & L^{BU}(n+1|n-1,n) + L^{UB}(n+1|n-1,n) = 0. \nonumber \\
\eea
Thus $\sum_n (|U_n|^2 + |B_n|^2)/2$ is conserved due to these energy transfers; the energy is exchanged among the velocity and magnetic fields.   Again following the same procedure as outlined in Appendix~\ref{appendix:fluid} we derive the shell-to-shell energy transfers formulas for $U2B$ transfers as
\begin{eqnarray}
S^{BU}(n+1|n|n-1) & = & b_{2} P^{BU}_{B}(n+1|n|n-1), \\
S^{BU}(n+1|n-1|n) & = & b_{1} P^{BU}_{B}(n+1|n-1|n), \\
S^{BU}(n|n+1|n-1) & = & b_{3}  P^{BU}_{B}(n|n+1|n-1), \\
S^{BU}(n|n-1|n+1) & = & b_{1} P^{BU}_{B}(n|n-1|n+1), \\
S^{BU}(n-1|n|n+1) & = & b_{2} P^{BU}_{B}(n-1|n|n+1), \\
S^{BU}(n-1|n+1|n) & = & b_{3} P^{BU}_{B}(n-1|n+1|n).
\end{eqnarray}
where $P^{BU}_{B}$ is defined in Eq.~(\ref{eq:P}).  Here the giver shell $X=U$, the receiver shell $Y=B$, and the mediator shell $Z=B$.  Also, by definition the energy gained by $B$ from $U$ is the energy lost by $U$ to $B$, hence
\begin{equation}
S^{UB}(n+1|n|n-1)=-S^{BU}(n|n+1|n-1). \label{eq:Smhd-balance-1} 
\end{equation}
Other $S^{UB}$'s can be written in the same manner.

Using the shell-to-shell energy transfers we can write down the energy flux, $\Pi^{U<}_{B>}$, which is the net energy transfers from the shells inside the $U$ sphere of radius $K$  to the shells of outside the $B$ sphere, which is
\begin{eqnarray}
\Pi_{B>}^{U<}(K) & = & \sum_{m\le K}  \sum_{n>K} \sum_{p}S^{BU}(n|m|p).
\label{eq:PiBg_Ul}  
\eea  
Similarly, the energy flux, $\Pi^{B<}_{U>}$, which is the net energy transfers from the shells inside the $B$ sphere of radius $K$  to the shells of outside the $U$ sphere, which is
\begin{eqnarray}
\Pi_{U>}^{B<}(K) & = & \sum_{m\le K}  \sum_{n>K} \sum_{p}S^{UB}(n|m|p).
\label{eq:PiBl_Ug}  
\eea  
We also have energy transfers from the $U$-shells inside the sphere to the $B$-shells inside the sphere, which is
\begin{eqnarray}
\Pi_{B<}^{U<}(K) & = & \sum_{m\le K}  \sum_{n \le K} \sum_{p}S^{BU}(n|m|p).
\label{eq:PiUl_Bl}  
\eea  
Similarly for the $U2B$ energy transfer formula for the shells outside the sphere is
\begin{eqnarray}
\Pi_{B>}^{U>}(K) & = & \sum_{m > K}  \sum_{n > K} \sum_{p}S^{BU}(n|m|p).
\label{eq:PiUg_Bg}  
\eea


\end{document}